\documentclass[twocolumn,showpacs,preprintnumbers,amsmath,amssymb]{revtex4}
\usepackage[colorlinks, linkcolor=blue, citecolor=blue]{hyperref}

\usepackage{graphicx}
\usepackage{epstopdf}
\usepackage{dcolumn}
\usepackage{bm}
\usepackage[textwidth = 2.5\linewidth]{todonotes}
\usepackage{xcolor}
\usepackage{soul}

\sethlcolor{green}

\begin{document}

	\title{Magnetic structure of the promising candidate for three-dimensional artificial spin ice: small angle neutron diffraction and micromagnetic simulations}

	\author{A.A.~Mistonov$^{1}$, I.S.~Dubitskiy$^{2}$, I.S.~Shishkin$^{2}$, N.A.~Grigoryeva$^1$, A.~Heinemann$^3$, N.A.~Sapoletova$^{4}$, G.A.~Valkovskiy$^{1}$, S.V.~Grigoriev$^{1,2}$}

	\affiliation{
	$^1$Department of Physics, Saint-Petersburg State University, 198504 Saint Petersburg, Russia \\
	$^2$Petersburg Nuclear Physics Institute, NRC “Kurchatov Institute”, 1, Orlova roscha mcr., Gatchina, Leningrad Region 188300, Russia \\
	$^3$Helmholz Zentrum Geesthacht, Geesthacht Germany \\
    $^4$Lomonosov Moscow State University, 119991 Moscow, Russia \\
	}

	\date{\today}
	\begin{abstract}
	
Geometrical frustration arised in spin ices leads to fascinating emergent physical properties. Nowadays there is a wide diversity of the artificial structures, mimicking spin ice at the nanoscale and demonstrating some new effects. Most of the nanoscaled spin ices are two dimensional. Ferromagnetic inverse opal-like structures (IOLS)  are among inspiring examples of the three-dimensional system exhibiting spin ice behaviour. However detailed examination of its properties is not straightforward. Experimental technique which is able to unambiguously recover magnetization distribution in 3D mesoscaled structures is lacking. In this work we used an approach based on complementary exploiting of small-angle neutron diffraction technique and micromagnetic simulations. External magnetic field was applied along three main directions of the IOLS mesostructure. Comparison of the calculated and measured data allowed us to determine IOLS magnetic state. The results are in good agreement with the spin ice model. Moreover influence of the demagnetizing field and vortex states on the magnetizing process were revealed. Additionally, we speculate that this approach can be also applied to other 3D magnetic mesostructures.

	\end{abstract}
	\maketitle

%%%%%%%%%%%%%%%%%%%%%%%%%%%%%%%%%%%%%%%%%%%%%%%%%%%%%%%%%%%%

%%%%%%%%%%%%%%%%%%%%%   INTRODUCTION SECTION   %%%%%%%%%%%%%%%%%%%%%%%%

\section{Introduction}
\label{sec:introduction}

Spatially ordered ferromagnetic nanosystems attract significant attention, since they can mimic natural objects (e.g. pyrochlore spin ices), but at the same time bring some new physics. 3D nanostructures are of considerable interest because of possible qualitatively new properties~\cite{fernandez2017three}. The promising example is inverse opal-like structure (IOLS) based on ferromagnetic metal~\cite{grigoriev2009structural, grigoryeva2011magnetic}. It can be considered as the three-dimensional continuous network, consisting of cubical and tetrahedral shaped nanoscaled elements with concave faces, connected by long narrow links (``legs'') and arranged into face-centered cubic (fcc) lattice. Non-trivial geometry of the IOLS causes a complicated distribution of the magnetization, which cannot be unravelled by conventional techniques. It was shown in~\cite{mistonov2013three,dubitskiy2015spin}, that magnetization distribution in IOLS is almost fully determined by the direction of the magnetic moments in ``legs'' which are arranged along four $\langle 111\rangle$-type axes of the fcc IOLS mesostructure. A strong shape anisotropy of the ``legs'' leads to an uniform distribution of the magnetization inside them. In other words the ``legs'' appear to be of Ising-type in a wide range of external fields. The magnetic structure of the IOLS is mostly governed by two ``forces'': external magnetic field \textbf{H} and the ice rule. In our case, the ice rule states, that the number of ``legs'' with the magnetic moments pointing inward and outward each nanoobject (cube or tetrahedron) should be equal.

Therefore IOLS can be considered to belong to the family of nanostructured artificial spin ices (ASI)~\cite{nisoli2013colloquium, wang2006artificial}. Initially ASI were designed to mimic atomic spin ice systems firstly discovered in pyrochlore lattices \cite{bramwell2001spin, den2000dipolar}. Various realizations of ASI demonstrate emergent physical properties that can be even richer than those of natural materials~\cite{nisoli2018topology}. However, two-dimensional ASI with coplanar arrangement of magnetic moments differ from the conventional atomic spin ice systems. On the contrary, the inverse opals are three-dimensional and can be considered as a more direct mesoscale analogue of pyrochlore spin ice. Additionally, in recent years~\cite{lee2007iron, hsueh2011nanoporous,Eslami2014,mamica2016magnetoferritin,Utke2008} the number of three-dimensional nano- and mesoscaled systems increases drastically~\cite{fernandez2017three} owing to development of the synthesis techniques. This underlines the importance of determination of the three-dimensional magnetic structure. Nevertheless experimental investigation of the magnetization distribution inside such systems is still challenging.

%In current paper we apply the approach, which will be described below, to resolve the problem.  An interplay between  external magnetic field and the "ice rule" leads to different magnetization reversal scenarios depending on the direction of the field.

%There are three types of the interplay: \textit{``competition''}, when the configuration of magnetic moments in ``legs'' corresponded to minimum Zeeman energy leads to the ``ice rule'' violation;  \textit{``independence''}, when the ``ice rule'' is fulfilled at any field value (below ``legs'' anisotropy breaking), but field removes the degeneracy; \textit{``cooperation''}, when ``ice rule'' and field tend to lead the system to the same configuration. These situations almost totally cover all possible scenarios of magnetic structure evolution. However there is one more case, we have to mention. It can be referred as \textit{``super-independence''}, when the ``ice rule'' is valid below ``legs'' shape anisotropy breaking and furthermore degeneracy is not fully removed by the field. It leads to the partial loss of magnetic long-range order~\cite{mistonov2015ice}. Here we focus on ordered states.

It is possible to determine three-dimensional magnetic structure at atomic scale (e.g. by neutron diffraction) as well as at microscale (e.g. using tomography), while for nano and mesoscale (hundreds of nanometers) this is non-trivial. Conventional methods are limited to simple forms of samples~\cite{Wernsdorfer1996nucl,Biziere2013imaging}, small sample sizes~\cite{yu2010real,Zhao03052016} or modest spatial resolution~\cite{manke2010three}. In order to understand fine features of the magnetic structure we suggest to exploit simultaneously small-angle neutron diffraction and micromagnetic simulations.

Small-angle neutron scattering (SANS) is a powerful technique for the mesoscopic magnetic structure studies~\cite{fitzsimmons2014neutron, grutter2017complex, grigoriev2010nanostructures, gunther2014magnetic}. In principle it allows one to overcome the above mentioned limitations. However, in most cases the interpretation of the SANS data is not simple and can be inconclusive. Thus, other techniques should be involved, which can provide either two-dimensional surface (magnetic force microscopy (MFM)~\cite{hartmann1999magnetic}, Lorenz microscopy~\cite{de20012}) or integral (superconducting quantum interference device (SQUID)-magnetometry) information about the sample. Another approach is to use micromagnetic calculations, which are able to replace or to complement actual measurements~\cite{bertotti1998hysteresis}.

In order to interpret SANS data unambiguously one has to calculate Fourier transform of the magnetization distribution in the sample under study~\cite{gunther2014magnetic, ott2013numerical}. In most cases the magnetization distribution is assumed to be uniform ~\cite{michels2014magnetic}. This assumption is reasonable for small or saturated nanoparticles however it is not acceptable for more complex systems. In the latter case one has to determine magnetization distribution in the system in order to find further the Fourier transform of this distribution. In doing so, the most rigorous and straightforward way    is to solve micromagnetic equations~\cite{gilbert1955lagrangian, bertotti1998hysteresis, aharoni2000introduction}. The apparent advantage of the micromagnetic calculations is that no adjustable parameters have to be used. The magnetization distribution depends on well known parameters of the material and its geometric arrangement. We used the scheme described above, in order to calculate the Fourier transform of the magnetization distribution in the IOLS mesostructure unit cell and its evolution during the magnetizing process.

In this paper we suggest to consider three scenarios of the interplay between the external magnetic field and the ice rule and reveal magnetic structure using SANS and micromagnetic simulations in pair. This combination was successfully applied to two-phase bulk ferromagnets~\cite{michels2014magnetic, michels2014micromagnetic, erokhin2012micromagnetic}. Additionally, simulation results were recently reported for the single nanowire~\cite{vivas2017small}.

We assume, that the proposed approach can be widely used for the investigation of magnetic nanostructures and mesoscopic three-dimensional periodical systems like segmented nanowires~\cite{Sergelius2017}, nanosprings~\cite{Nam2017}, gyroids~\cite{hsueh2011nanoporous}, mesocrystals~\cite{Sturm2016}.

The paper is organized in the following way. Section~\ref{sec:samples} gives the essence of the sample preparation procedure. Section~\ref{sec:SANS} describes the small-angle neutron scattering experiment and some details of data treatment. Section~\ref{sec:MM_calculations} gives the description of the micromagnetic simulation procedure. In Section~\ref{sec:results} the results of the SANS experiment and micromagnetic calculations are presented along with discussion. Section~\ref{sec:conclusion} gives concluding remarks.

%%%%%%%%%%%%%%%%%%%%%   SAMPLES SECTION   %%%%%%%%%%%%%%%%%%%%%%%%
\section{Samples}
\label{sec:samples}

The inverse opal-like Co structure was prepared by electrodeposition of cobalt inside the voids of the artificial opal. The artificial opal template was synthesized by electric-field-assisted vertical deposition of polystyrene microspheres (D~=~620~nm; RSD~$<$~10\%)~\cite{napolskii2010} on Si (100) wafer PVD-coated with a 200~nm thick gold layer. The cobalt electrocrystallization into the opal voids was carried out in three-electrode cell~\cite{sapoletova2010controlled} from the electrolyte containing 0.2M~CoSO$_4$ + 0.5M~Na$_2$SO$_4$ + 0.3M~H$_3$BO$_3$ + 3.5M~C$_2$H$_5$OH at deposition potential -0.8~V versus Ag/AgCl reference electrode at room temperature. The area of the sample was 1~cm$^2$. In order to obtain the free-standing metallic structure on the support, the polystyrene microspheres were dissolved in toluene for several hours.

\begin{figure}[h]
  \begin{minipage}{0.49\linewidth}
    \center{\includegraphics[width=1\linewidth]{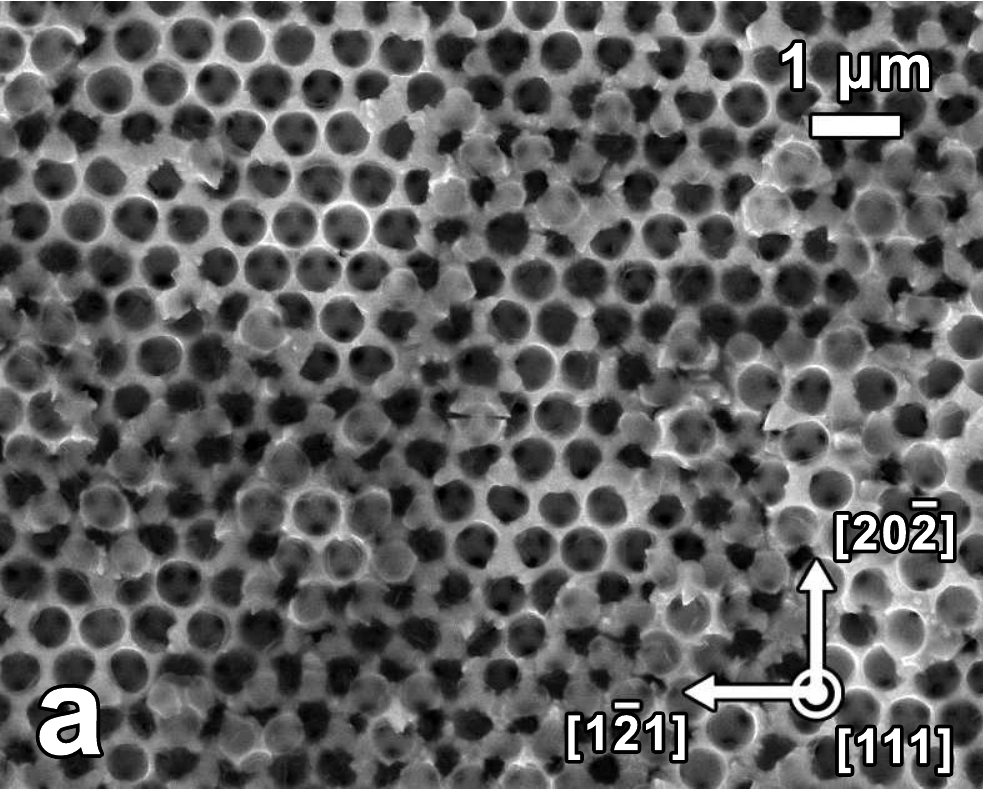}}
  \end{minipage}
  \begin{minipage}{0.49\linewidth}                 \center{\includegraphics[width=1\linewidth]{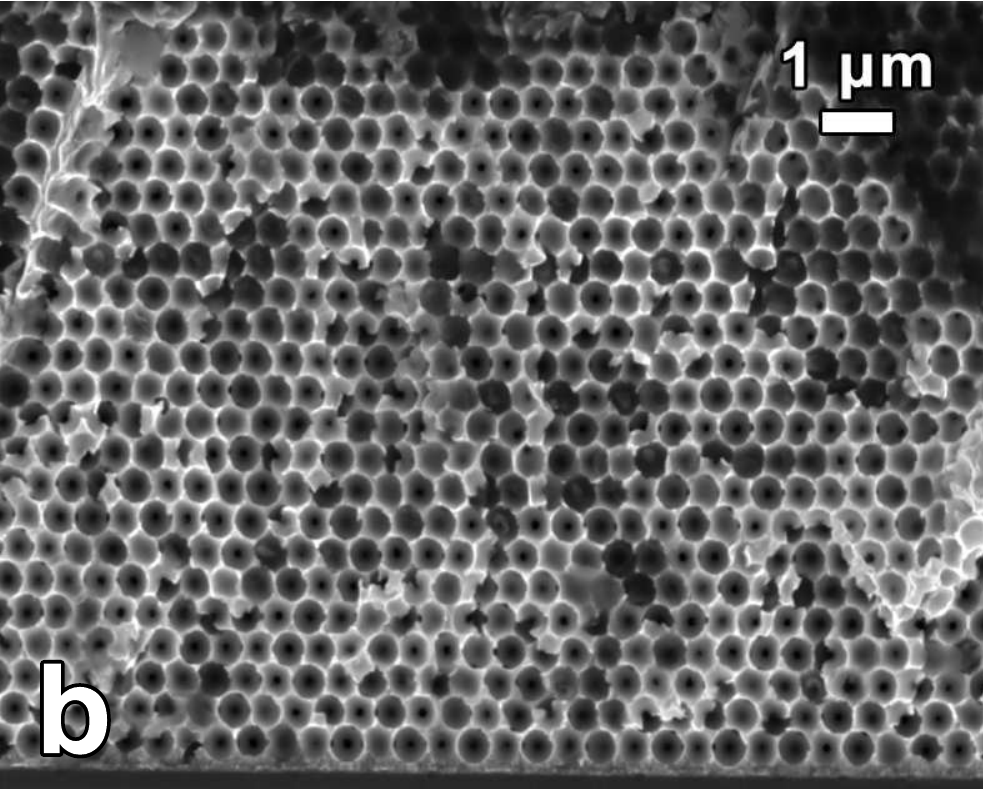}}
  \end{minipage}
 \caption{ (a) Top and (b) side views of the Co IOLS obtained by SEM.}
 \label{ris:Co_SEM}
\end{figure}

Scanning electron microscopy (SEM) images of the Co IOLS are given in Fig.~\ref{ris:Co_SEM}. One can see, that the surface represents a hexagonal closepacked layer of spherical voids with the lattice constant of about 600~nm. An average lateral size of the structural domain exceeds 100~$\mu$m. Desorientation of different domains is less, than 3$^{\circ}$. According to the SEM data thickness of the Co IOLS is 12~$\mu$m. With the help of the SEM images one can estimate the degree of polystyrene microspheres deformation (sintering) which determines ``legs'' aspect ratio~\cite{dubitskiy2015spin, dubitskiy2017dependence}. These quantity is defined as $r/r'-1$ expressed in percents, where $r$ --- microsphere radius, $r'$  --- half of the distance between adjacent microspheres centers. It is equal to 2\% for the considered sample.

Spatial ordering of the sample on the mesoscale was investigated by microradian x-ray diffraction technique. Details of data acquisition and treatment can be found elsewhere~\cite{grigoriev2009structural,grigoryeva2011magnetic,chumakova2014periodic, dubitskiy2017study}. It was shown, that the sample possesses fcc symmetry with lattice constant $a_{0}$~=~845$\pm$10~nm.

The method of sample fabrication allows one to synthesize structures  with orientation of the [111] crystallographic axis of the IOLS fcc mesostructure perpendicular to the substrate (here and further we deal only with crystallographic axes related to IOLS mesostructure, not to the cobalt atomic structure). The direction of drying of the aqueous suspension of polystyrene microspheres (or meniscus moving) corresponds to the [20$\bar{2}$] axis. Thus, the orientation of the opal-like crystal is already established at the stage of synthesis. The main axes those determine the sample orientation are shown in Fig.~\ref{ris:Co_SEM}.

\begin{figure}[h]
	\begin{minipage}{0.99\linewidth}
		\center{\includegraphics[width=1\linewidth]{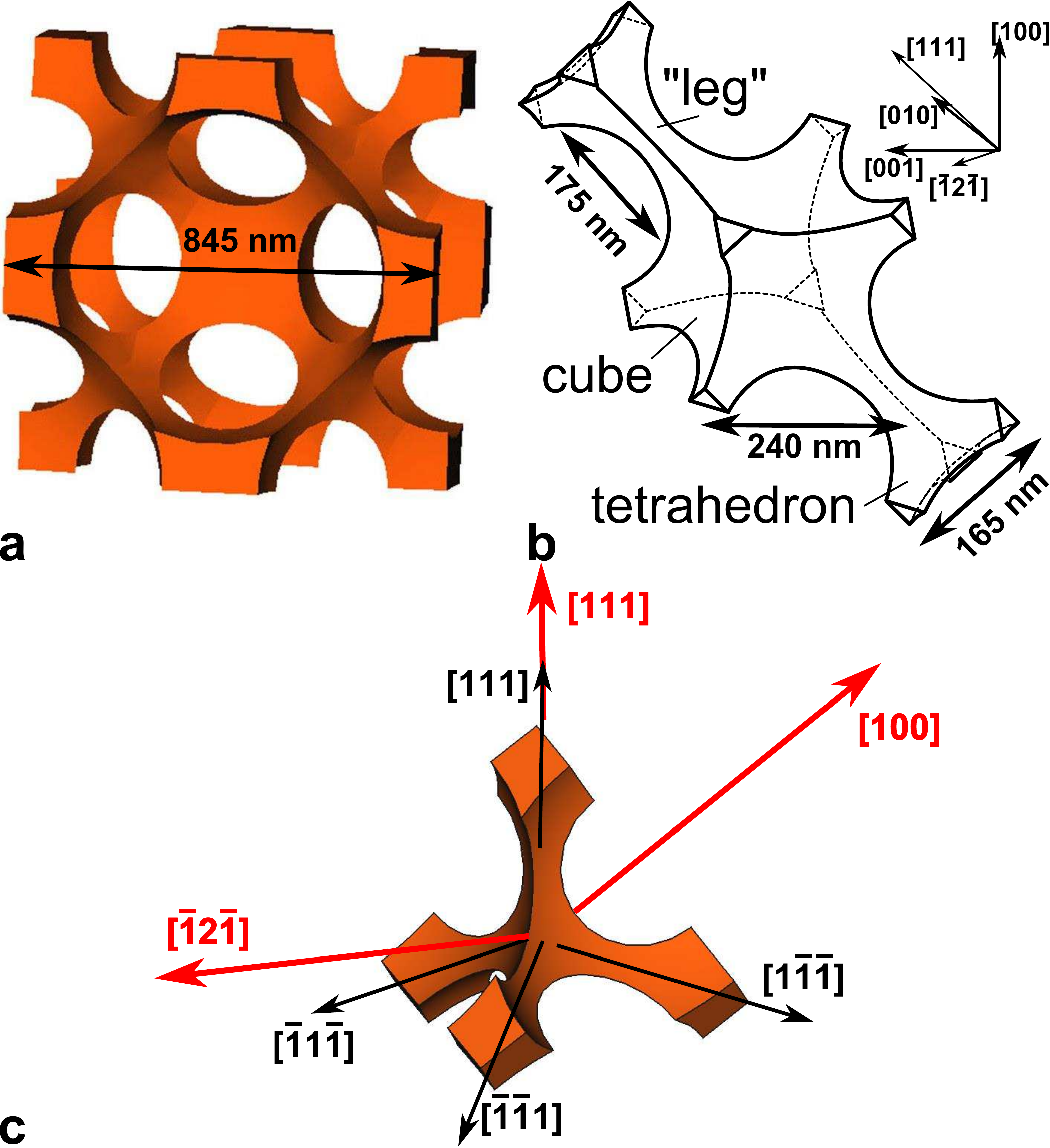}}
	\end{minipage}
% 	\begin{minipage}{0.49\linewidth}
% 		\center{\includegraphics[width=1\linewidth]{Primitive_cell.eps}}
% 	\end{minipage}
	\caption{(Color online) (a) Unit cell and  (b) primitive cell of the inverse opal-like structure. (c) Tetrahedron with four directions of $\langle$111$\rangle$-type axes and three directions, along which the field was applied (for better visibility the sintering degree equals to 10\%). The sizes of ``leg'', cube and tetrahedron edges are also shown.}
	\label{ris:Basic_element}
\end{figure}

The appearance of the structural elements of the IOLS can be easily understood. We described them elsewhere~\cite{mistonov2013three,mistonov2015ice,dubitskiy2015spin,dubitskiy2017dependence} but outline briefly here. The tetrahedral and octahedral voids of the fcc structure transform into two types of nanoobjects during the inversion process. We denote them quasitetrahedra and quasicubes, but further ``quasi'' is omitted. In Fig.~\ref{ris:Basic_element} the unit cell (a) and the primitive cell (b) of the IOLS are presented. The primitive cell consists of three parts: a tetrahedron, a cube and another tetrahedron. They are connected to each other by ``legs'' along one of four $\langle 111 \rangle$-type axes. The surfaces of the cube and tetrahedra are concave resembling the voids between the spheres. For IOLS with the period of 845~nm one can estimate that the cube has edge of about 240~nm, while tetrahedra have edges of about 165~nm . The connecting ``legs'' have a length of about 175~nm.

%%%%%%%%%%%%%%%%%%%%%   METHODS SECTION   %%%%%%%%%%%%%%%%%%%%%%%%

\section{Methods and data treatment}
\label{sec:methods}

%%%%%%%%%%%%%%%%%%%%%   SANS SUBSECTION   %%%%%%%%%%%%%%%%%%%%%%%%

\subsection{Small-angle neutron diffraction}
\label{sec:SANS}

\begin{figure*}[!htbp]
	\begin{minipage}{0.99\linewidth}
		\center{\includegraphics[width=1\linewidth]{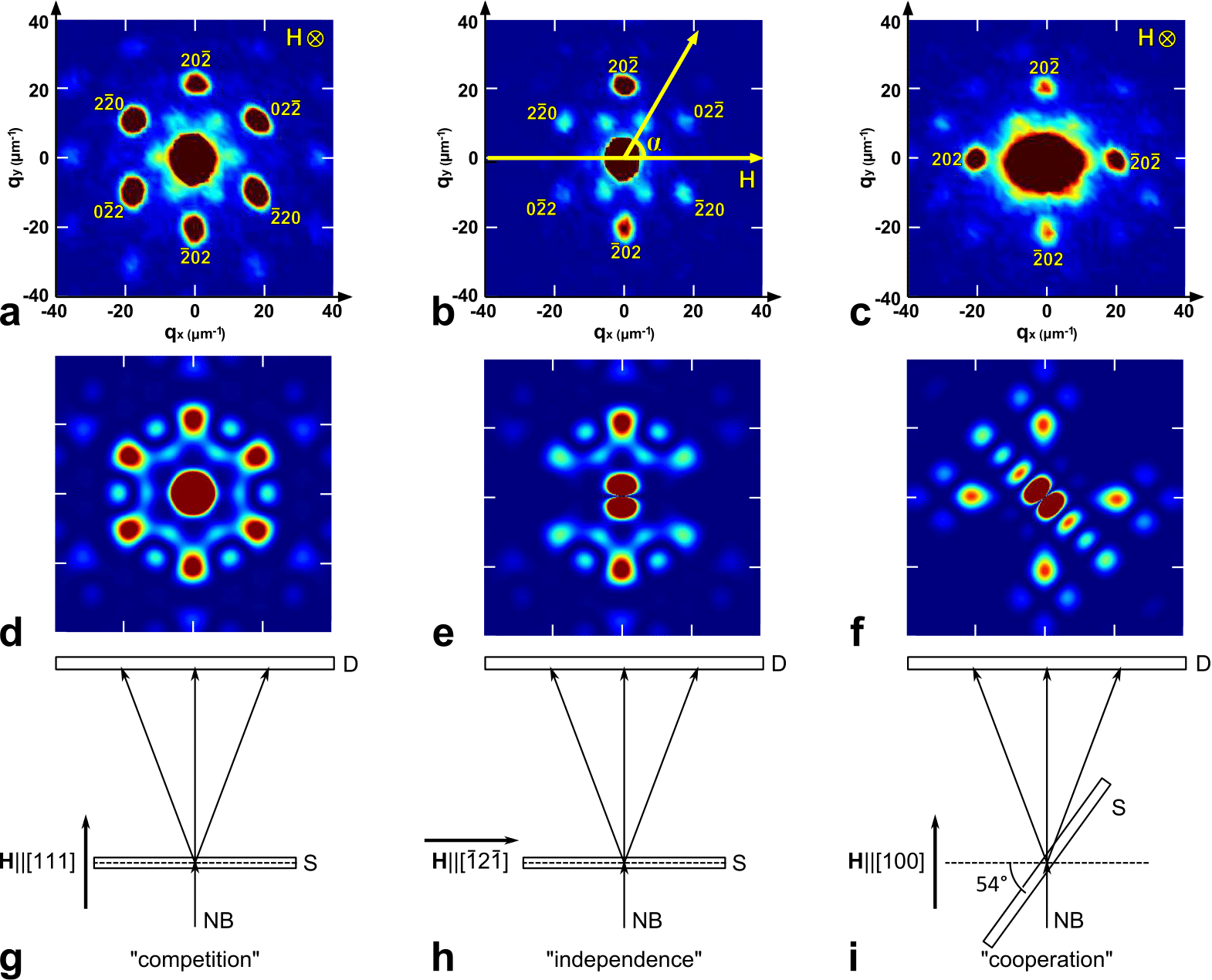}}
	\end{minipage}
	\caption{(Color online) Experimental small-angle neutron diffraction patterns for Co IOLS (top row); calculated Fourier transform $\widetilde{\textbf{M}}_{\perp}(\textbf{q})$  of the magnetization distribution at the correponding cross section of the reciprocal space (middle row); top view of the experimental geometries (bottom row). Intensity scale is linear. External magnetic field $\textbf{H}$ is applied along [111] (first column), $[\bar{1}2\bar{1}]$ (second column) and [100] (third column) directions.  NB -- neutron beam, S -- sample, D -- detector}
	\label{ris:exp_geometry}
\end{figure*}

SANS measurements were carried out with the instrument SANS\nobreakdash-1 (Heinz Maier Leibnitz Zentrum, Garching, Germany). Neutron beam with a mean wavelength of $\lambda$=1.7~nm and a wavelength spread of $\Delta \lambda/\lambda$~=~10\% was used. The Sample - Detector distance of 21.719~m was chosen with appropriate collimation to cover scattering vectors q from 0.005 to 0.06~nm$^{-1}$. Due to the large wavelength, the intensity of the incident beam was low enough to work without a beamstop. Thus the transmitted beam as well was recorded by the detector. Since the system ordered at the mesoscale small-angle scattering was actually small-angle diffraction. An external magnetic field up to 30~kOe was applied along the [111], $[\bar{1}2\bar{1}]$ and [100] crystallographic axes of the fcc mesostructure. In first two cases, the sample was irradiated by the beam along [111] axis, while in latter --- along [100] axis. All measurements were performed at room temperature. Geometrical schemes used in the experiment as well as recorded diffraction patterns are shown in Fig.~\ref{ris:exp_geometry}.

The intensity of non-polarized neutrons scattering consists of the following  contributions:
\begin{equation}
	I_{tot}=I_N+I_M,
    \label{eq:total_section}
\end{equation}
\begin{equation}
	I_N(\textbf{q})\propto{S_n(\textbf{q})|A_nF_n(\textbf{q})|^2},
    \label{eq:nuclear_section}
\end{equation}
\begin{equation}
	I_M(\textbf{q})\propto{S_m(\textbf{q})|A_m|^2|\widetilde{\textbf{M}}_{\perp}(\textbf{q})|^2},
    \label{eq:magnetic_section}
\end{equation}

where $I_N$ and $I_M$  are intensities of the nuclear and magnetic scattering respectively, \textbf{q} --- scattering vector, S$_n$, S$_m$ are nuclear and magnetic structure factors, $A_n$ and $A_m$ are amplitudes of nuclear and magnetic scattering, respectively, F$_n$ --- nuclear form-factor and $\widetilde{\textbf{M}}_{\perp}(\textbf{q})$ -- the magnetic form-factor component perpendicular to $\textbf{q}$. For the IOLS with the voids of radius R the F$_n$ is equal to that of the sphere of radius R~\cite{FeiginSvergun}, S$_n$ is the structure factor of the fcc lattice with the lattice constant a$_0$. $\widetilde{\textbf{M}}(\textbf{q})$  is the Fourier transform of the magnetization distribution in the IOLS unit cell and S$_m$ is determined by the correlation of the magnetization distributions between unit cells.

Both nuclear and magnetic scattering intensities contain the contribution from periodical structure (Bragg peaks) as well as from disordered elements (diffuse scattering). Since we are interested in magnetic ordering in the IOLS unit cell we should exclude diffuse scattering which is caused by magnetic domains.

The diffraction patterns presented in Fig.~\ref{ris:exp_geometry} were taken from fully magnetized samples therefore the intensity of the peaks contains both nuclear and magnetic contributions. The data treatment procedure is described elsewhere~\cite{grigoriev2009structural,grigoryeva2011magnetic}, here we summarize it shortly. In order to extract the magnetic contribution caused by periodical magnetic structure one has to subtract nuclear contribution as well as diffuse one. At first we integrated diffraction pattern azimuthally near the reflex in order to obtain I$_{tot}$(q)-dependence. Then we approximated this dependence by the sum of diffuse (squared Lorenz function) and Bragg (Gauss function) contributions and determined the integral intensity of the latter. Finally we subtracted the value, obtained for demagnetized sample (at the external magnetic field equalled to the coercive force H$_{\text{C}}$), containing mostly nuclear contribution, since magnetic scattering caused by disordered domains has been subtracted in the previous step. Thus, we obtained field dependencies of the Bragg peaks intensities.

It is well known that diffraction patterns can be obtained in the experiment only if neutron coherence length is much larger than the period of the structure~\cite{grigor2007two, petukhov2002high, grigoriev2010nanostructures}. Transverse coherence length of the neutron beam $l_{\text{tr}}$ determines minimal peak width. This length depends on angular size of the neutron source~\cite{benfield2004structure}. In our case it is equal to 3~$\mu$m . Longitudinal coherence length of the neutron beam $l_{\text{long}}$ can be found from the spectral width of the source $\Delta\lambda$~\cite{born1980principles}. For the experimental setup used in present work $l_{\text{long}}$ equals to 1~mm. Therefore total coherence volume can be estimated as $l_{\text{tr}}^2l_{\text{long}}$~=~9000~$\mu$m$^3$. It covers for about 25 IOLS unit cells. At the same time beam spot size is about 1~cm$^2$. As a result the diffraction patterns consist of the incoherent sum of the waves scattered by different parts of the sample. It limits the analysis of the structure factor contribution in terms of peaks shape and peaks width. Another option is to consider the integral peak intensity. In this case one has to take into account the magnetic form factor contribution to the peak intensity. This contribution appears to be the most significant and substantially varies with variation of the external field.

%\begin{figure}[h]
%	\begin{minipage}{0.33\linewidth}
%		\center{\includegraphics[width=1\linewidth]{Co_26_H_111_I_m.png}}
%	\end{minipage}
%	\begin{minipage}{0.33\linewidth}
%		\center{\includegraphics[width=1\linewidth]{Fig1b.png}}
%	\end{minipage}
%	\begin{minipage}{0.33\linewidth}
%		\center{\includegraphics[width=1\linewidth]{Fig1c.png}}
%	\end{minipage}
%	\caption{Diffraction patterns for the magnetized state of the Co IOLS, when the field applied along [111] (a), [100] (b) and $[\bar{1}2\bar{1}]$}
%	\label{ris:Diffraction-pattern}
%\end{figure}

%%%%%%%%%%%%%%%%%%%%%   MICROMAGNETIC CALCULATIONS SUBSECTION   %%%%%%%%%%%%%%%%%%%%%%%%

\subsection{Micromagnetic calculations}
\label{sec:MM_calculations}

Micromagetic simulations were carried out by means of Nmag \cite{fischbacher2007systematic, fischbacher2009parallel} modelling package provided by University of Southampton. Nmag implements time integration of Landau-Lifshitz-Gilbert equation using ``Finite Element Method/Boundary Element Method''~\cite{fredkin1990hybrid}. Boundary element matrix was approximated by hierarchical matrices~\cite{hackbusch1999sparse}. We used the finite element discretization scheme since IOLS unit cell contains a lot of curved facets.

We simulated magnetization distribution in the inverse opal unit cell under the field applied along [111], [100] and $[\bar{1}2\bar{1}]$ crystallographic directions of IOLS mesostructure (Fig.~\ref{ris:Basic_element}) by using the same approach, as in Ref.~\cite{dubitskiy2015spin}. The degree of sintering was chosen to be 2\% according to the experimental data.

We used the following material parameters of bulk Co: the exchange coupling constant $A=3\cdot10^{-11}$ J/m and the saturation magnetization $M_S$~=~1.4$\times$10$^6$~A/m~\cite{abo2013definition, han2009influence}. Exchange, demagnetization and Zeeman energies were taken into account.

Wide-angle powder x-ray diffraction experiments showed that Co-based IOLS consist of polycrystalline hcp cobalt with the grain size less than 30~nm~\cite{grigoryeva2011magnetic}. No distinct texture was found. The grains play a significant role in the magnetic behaviour of bulk polycrystalline ferromagnets~\cite{michels2014micromagnetic, sepehri2013high}. Nevertheless, the magnetic state of nanostructures fabricated from polycrystalline cobalt is mostly determined by the shape anisotropy and magnetocrystalline term is often neglected~\cite{castan2014magnetic, rodriguez2014high, phatak2014visualization, fernandez2009magnetization, pathak2014experimental}. It should be noted that hysteresis loops measured for polycrystalline fcc Ni- and hcp Co-based IOLS with the same structural period do not exhibit any qualitative difference~\cite{dubitskiy2015spin} although magnetocrystalline anisotropy of fcc nickel phase is significantly smaller than that of hcp cobalt. Moreover, numerical simulation for Ni- and Co-based IOLS shows a qualitative agreement with the experiment in spite of neglecting the magnetocrystalline anisotropy contribution~\cite{dubitskiy2015spin}. It implies that magnetocrystalline anisotropy is not the major energy term which determines the inverse opal magnetic structure. Furthermore taking polycrystallinity into account gives rise to adjustable parameters (grain shape, size, etc.) which cannot be easily obtained from the experiment but can probably affect numerical simulations~\cite{erokhin2017optimization}. In order to not introduce adjustable parameters in our model and restrict ourselves to IOLS geometrical characteristics measured in the experiment, we neglected magnetocrystalline anisotropy.

The linear size of the finite element (tetrahedron) was chosen to be smaller than the exchange length $l_{\text{ex}}=\sqrt{2A/{\mu_0M^2_S}}=4.9$~nm. In our model open boundary conditions (OBC) were applied instead of periodic boundary conditions (PBC), since PBC is not expected to affect significantly the results of the simulation due to the large size of the unit cell and its high porosity, while require very large computational resources. It is necessary to take into account demagnetizing field related to the sample shape (thin film). If the field is applied perpendicular to the sample surface (along [111] direction) then the ``effective film'' model introduced in~\cite{dubitskiy2015spin} is sufficient to obtain  quantitative agreement with the experiment. However this model is not applicable to two other cases of the magnetic field directions. So we used the so-called macrogeometry approach~\cite{fangohr2009new} instead. We considered the array of 18 copies of IOLS unit cell arranged in (111) plane in order to simulate thin film shape. It was found out that the calculation results do not alter with increasing the number of copies. Full 50-points hysteresis curve simulation takes about month and a half on the eight node SMP machine. The typical value of RAM used is 60 GB.

In contrast to~\cite{dubitskiy2015spin}, where total magnetization was compared with the  experimental results of SQUID-magnetometry, here we extracted the magnetic form factor $\widetilde{\textbf{M}}(\textbf{q})$ of the IOLS unit cell. It involved numerous calculations of the Fourier transform of the magnetization distribution $\textbf{M}(\textbf{r})$. Namely, we calculated the Fourier transform of $\textbf{M}(\textbf{r})$ for all values of the external magnetic fields used in the SANS experiment:

\begin{equation}
	\widetilde{\textbf{M}}(\textbf{q})=\int\limits_{\text{unit~cell}}\textbf{M}(\textbf{r})e^{i\textbf{qr}}d\textbf{r}
\end{equation}

It should be noted, that integration is taken over the whole unit cell, which is of submicron size and contains four primitive cells; thus, the obtained results contain information about the different primitive cells.

Due to local defects of the IOLS mesostructure, the ``legs'' reversal fields and hence  the magnetic form factor values may have some dispersion~\cite{shen2012dynamics}. At a first approximation the calculated magnetic form factor of the ideal structure can be considered as the average value of the real one.

The form factor component $\widetilde{\textbf{M}}_{\perp}(\textbf{q})$ which is perpendicular to $\widehat{\textbf{q}}=\textbf{q}/q$ was calculated for the field values up to 30~kOe according to~\cite{squires2012introduction}:

\begin{equation}
	|\widetilde{\textbf{M}}_{\perp}(\textbf{q})|^2=|\widehat{\textbf{q}}\times \widetilde{\textbf{M}}(\textbf{q}) \times \widehat{\textbf{q}}|^2
\end{equation}
It provided the field dependence of the $\widetilde{\textbf{M}}_{\perp}(\textbf{q})$.

%%%%%%%%%%%%%%%%%%%%%   RESULTS AND DISCUSSION SECTION   %%%%%%%%%%%%%%%%%%%%%%%%

\section{Results and discussion}
\label{sec:results}

\begin{figure}[h]
	\begin{minipage}{0.90\linewidth}
		\center{\includegraphics[width=1\linewidth]{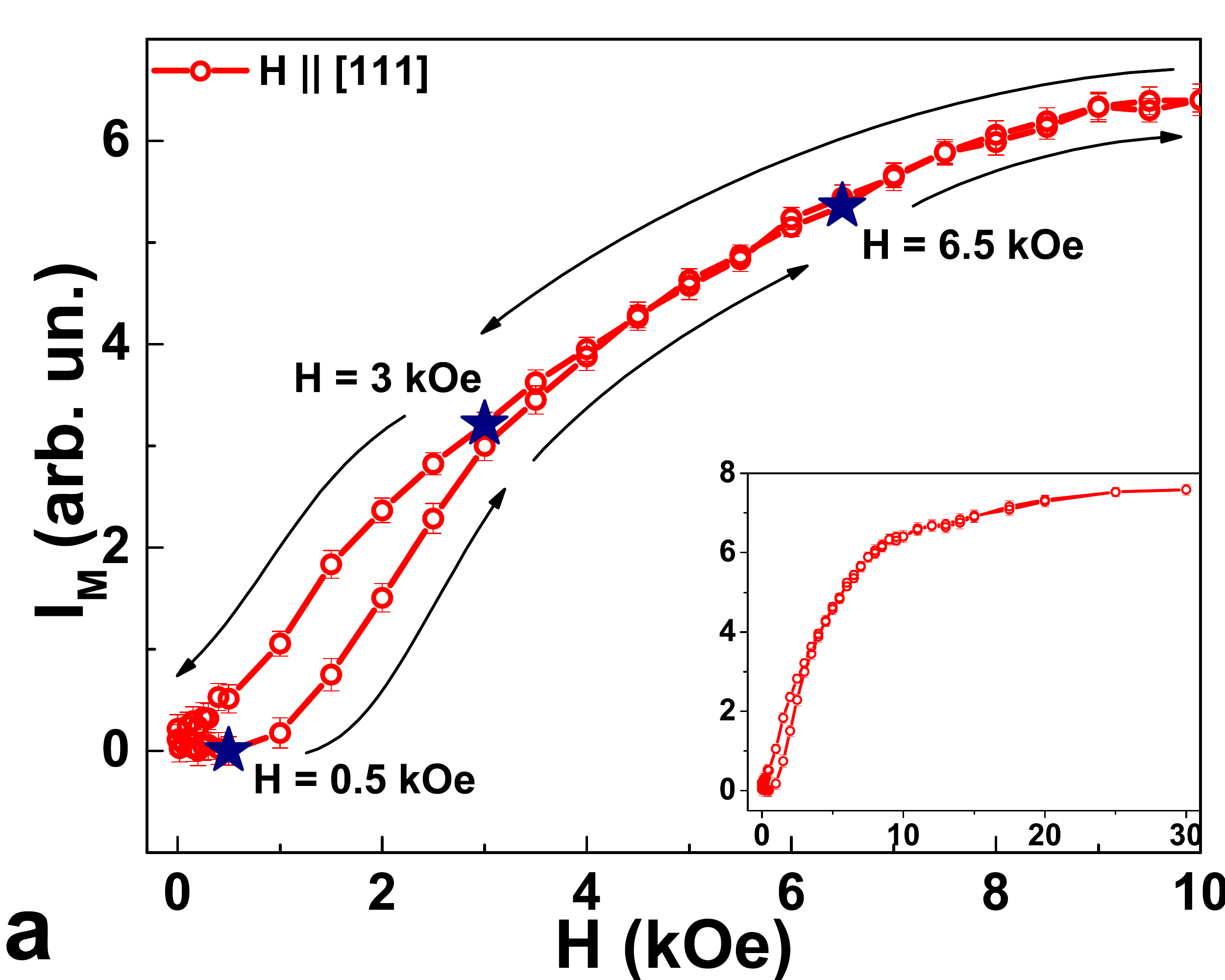}}
	\end{minipage}
	\begin{minipage}{0.90\linewidth}
		\center{\includegraphics[width=1\linewidth]{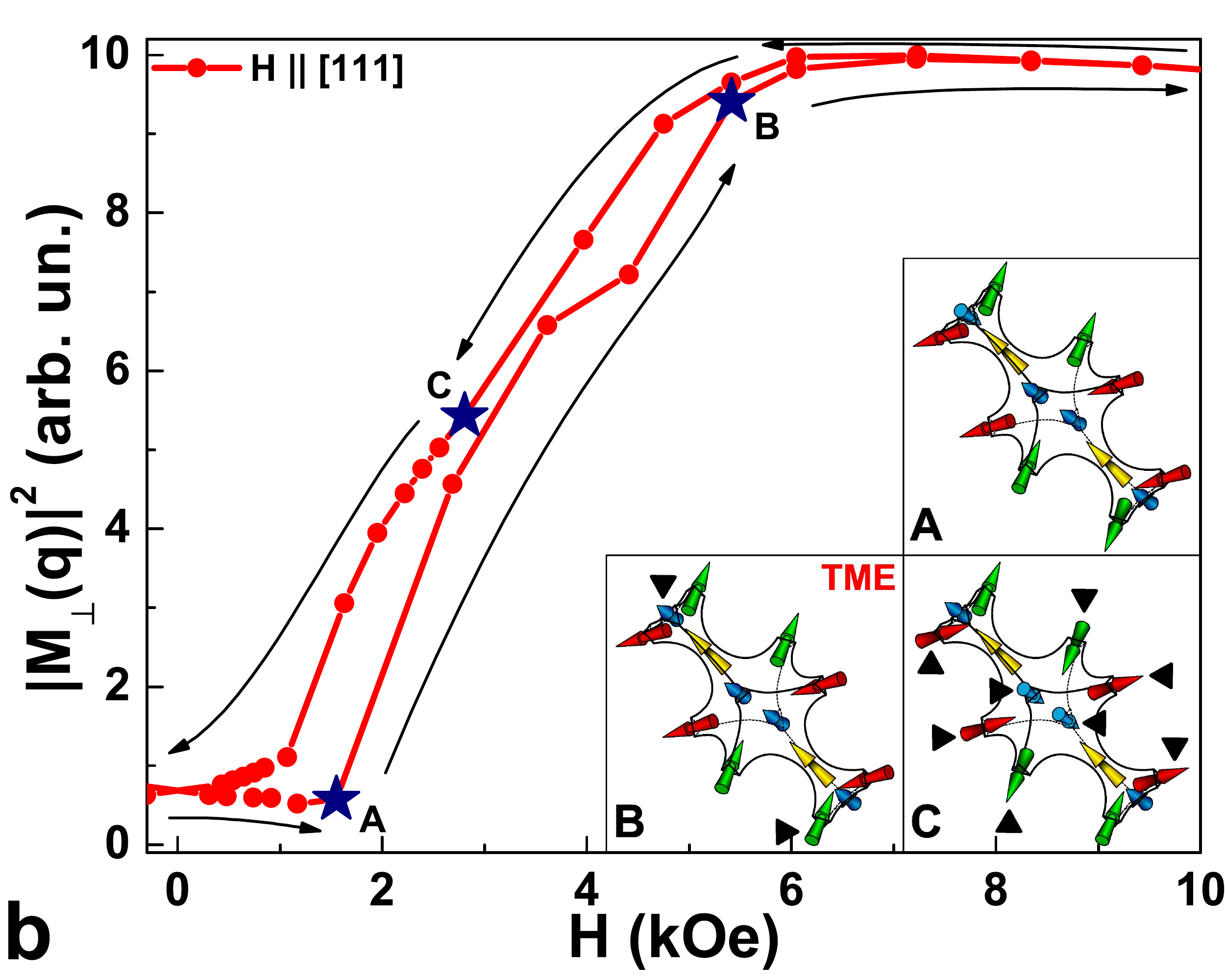}}
	\end{minipage}
	\caption{(Color online) Field dependence of (a) experimental neutron magnetic scattering intensity I$_M$ in the 202-type Bragg reflex positions for Co IOLS when the field is applied along [111] axis and (b) the value of the corresponding calculated magnetic form-factor. Solid lines are guides to the eye. Arrows show the sequence of the field values. For the selected parts of the curve, magnetic states of the primitive cell are shown in the insets. Black triangles in the insets mark the magnetic moments those changed the orientation with respect to the configuration at the previous value of the magnetic field.}
	\label{ris:H_111}
\end{figure}

\begin{figure}[!htbp]
	\begin{minipage}{0.75\linewidth}
		\center{\includegraphics[width=1\linewidth]{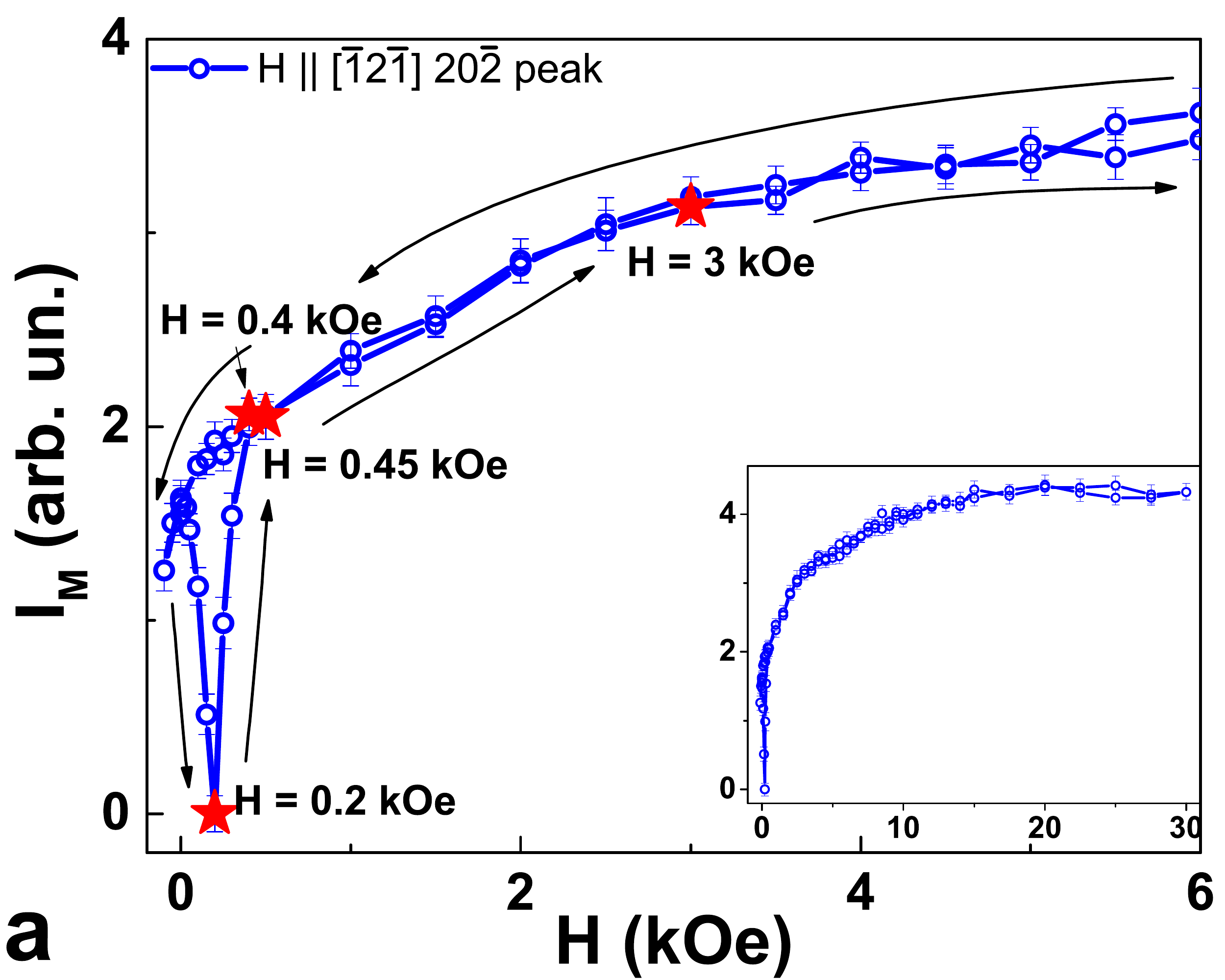}}
	\end{minipage}
	\begin{minipage}{0.75\linewidth}
		\center{\includegraphics[width=1\linewidth]{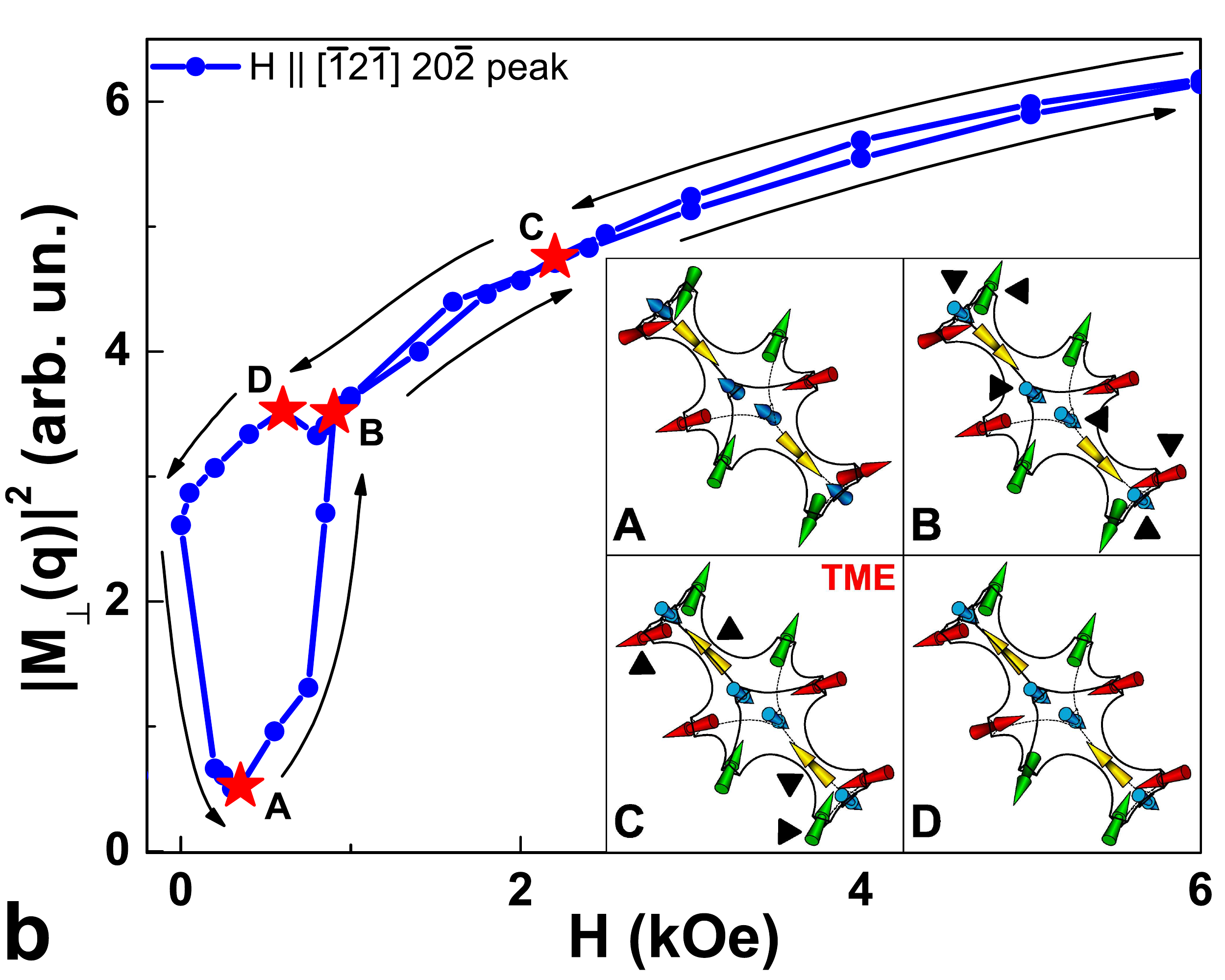}}
	\end{minipage}
	\begin{minipage}{0.75\linewidth}
		\center{\includegraphics[width=1\linewidth]{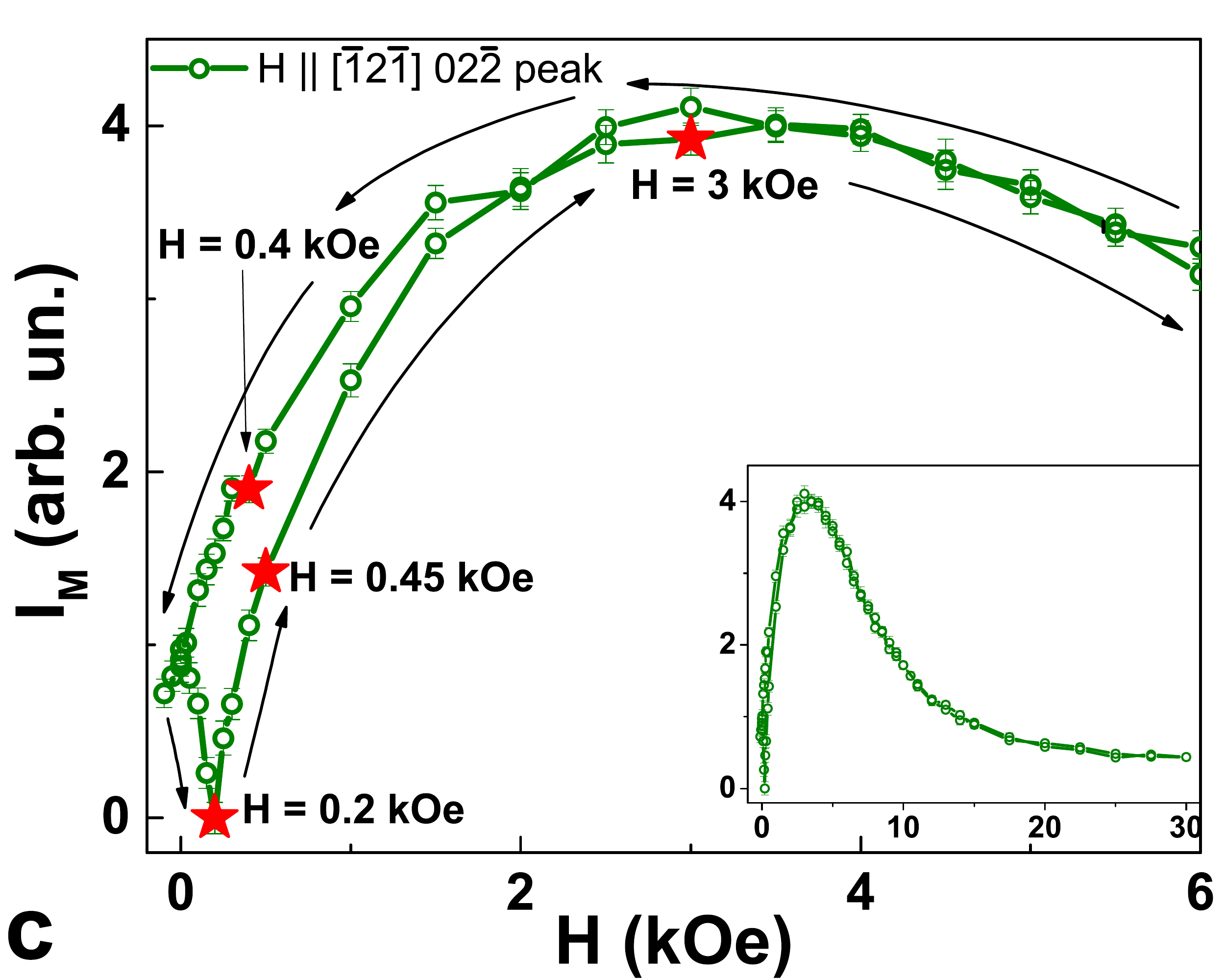}}
	\end{minipage}
	\begin{minipage}{0.75\linewidth}
		\center{\includegraphics[width=1\linewidth]{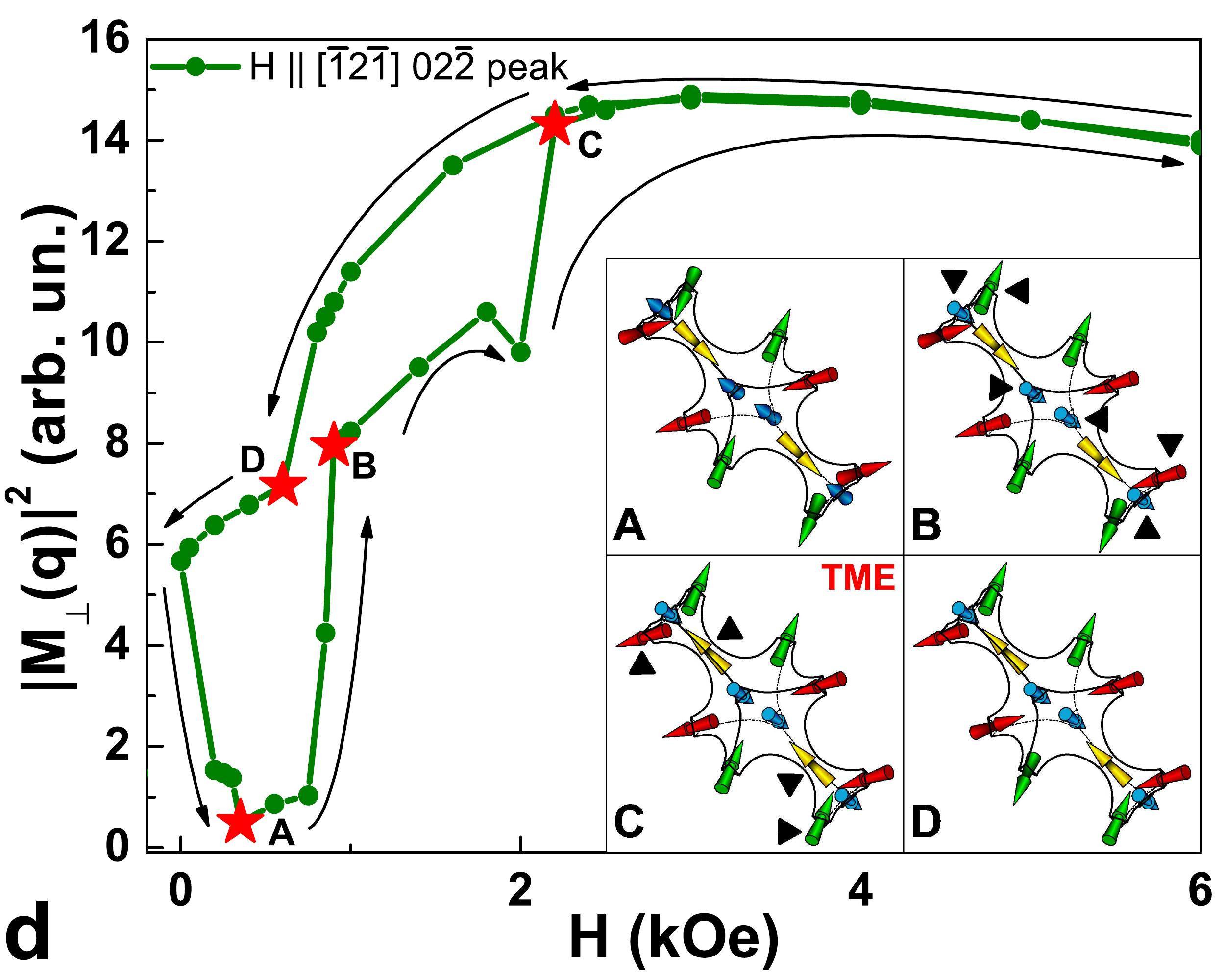}}
	\end{minipage}
	\caption{(Color online) Experimental field dependence of neutron magnetic scattering intensity I$_M$ for Co IOLS when the field is applied along [$\bar{1}2\bar{1}$] axis for (a) 20$\bar{2}$ and (c) 02$\bar{2}$ Bragg reflexes. The value of the calculated magnetic form-factor for (b) 20$\bar{2}$ and (d) 02$\bar{2}$ reflexes as a function of the external field. Solid lines are guides to the eye. Arrows show the sequence of the field values. For selected parts of the curves, magnetic states of the primitive cell are shown in the insets. Black triangles in the insets note the magnetic moments those changed the orientation with respect to the configuration at the previous value of the magnetic field.}
	\label{ris:H_121}
\end{figure}

\begin{figure}[!htbp]
	\begin{minipage}{0.90\linewidth}
		\center{\includegraphics[width=1\linewidth]{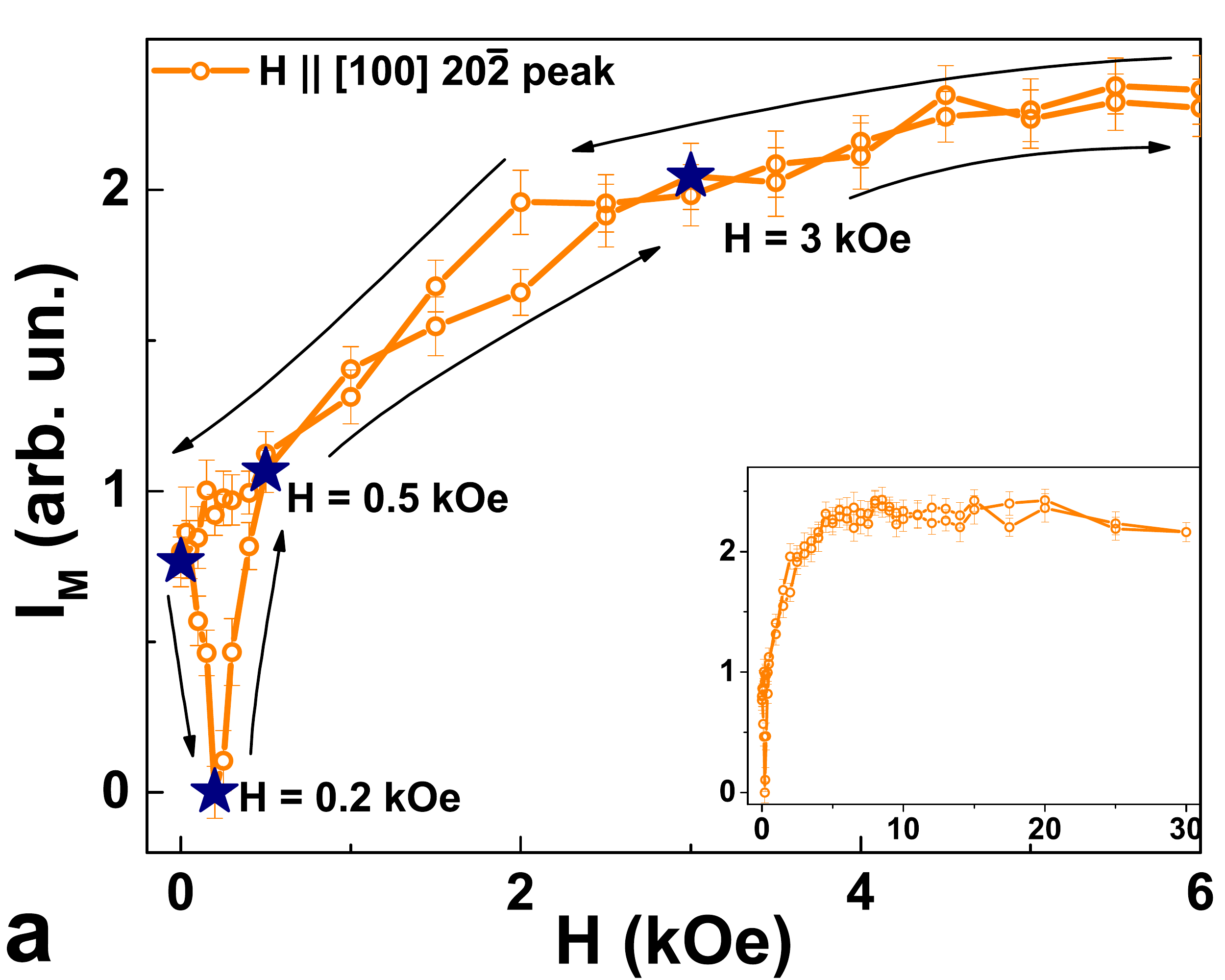}}
	\end{minipage}
	\begin{minipage}{0.90\linewidth}
		\center{\includegraphics[width=1\linewidth]{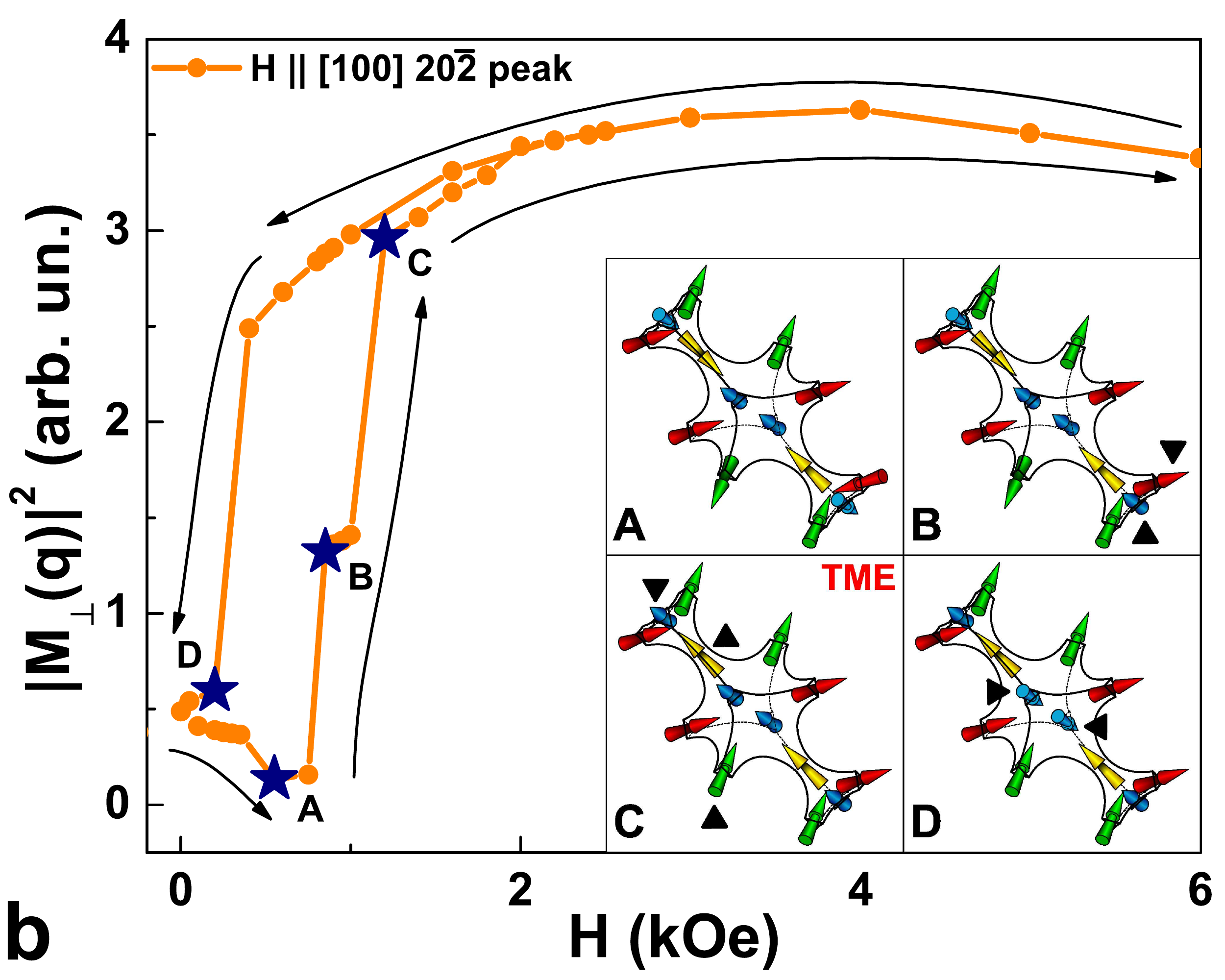}}
	\end{minipage}
	%	\begin{minipage}{0.90\linewidth}
	%		\center{\includegraphics[width=1\linewidth]{H_100_0}}
	%	\end{minipage}
	%	\begin{minipage}{0.90\linewidth}
	%		\center{\includegraphics[width=1\linewidth]{H_100_0_zoom}}
	%	\end{minipage}
	\caption{(Color online) Field dependence of (a) experimental neutron magnetic scattering intensity I$_M$ for Co IOLS when the field is applied along [100] axis and (b) the value of the corresponding calculated magnetic form-factor. Solid lines are guides to the eye. Arrows show the sequence of the field values. Black triangles in the insets note the vectors, changed the direction. For selected parts of the curve, magnetic states of the primitive cell are shown in the insets.}
	\label{ris:H_100}
\end{figure}

In this section we interpret the results of the neutron diffraction experiment by assigning magnetic states, obtained in micromagnetic simulations to the particular points of the experimental field dependencies. The states are shown in the insets in Figs.~\ref{ris:H_111}-\ref{ris:H_100} and corresponding points are marked by the same letters.

% \hly{Similarity in information which we extract from experimental data and simulations (see Sec.}~\ref{sec:methods}\hly{) allows us to compare magnetic part of the experimental Bragg peaks intensity and the value of the calculated magnetic form factor taken at the points of reciprocal space, where Bragg peaks should be located. Field dependence of these quantities has been considered.}

Firstly it is noteworthy to compare the two-dimensional experimental SANS patterns and the simulation results (Fig.~\ref{ris:exp_geometry}). One can observe the hexagonal arrangement of the Bragg reflexes in Fig.~\ref{ris:exp_geometry}a,b and four-fold symmetry in Fig.~\ref{ris:exp_geometry}c which is caused by the symmetry of the IOLS and the geometry of the experiment. The brightest well-resolved maxima in all cases correspond to diffraction at the crystallographic planes of \{202\}-family. The respective scattering vector is $q_{202}$~=~20.9~$\pm ~0.3~\mu$m$^{-1}$, i.e. the lattice constant of the fcc structure unit cell is a$_0$~=~840~$\pm$10~nm.

Although we did not take the structure factor into account directly the calculated patterns (being cross-sections of the magnetic form factor spatial distribution) are very similar to the corresponding experimental ones, those represent the intensity distribution (i.e. the product of the form factor and the structural one) [Fig.~\ref{ris:exp_geometry}d-f]. It means, that even four primitive cells stacked into the unit cell are sufficient to produce structural motif of the IOLS.

However, one can observe the difference between the experimental and calculated pictures. For instance, the central part of the diffraction patterns contains mainly the intensity of the direct beam, while in the calculated maps at the same place the magnetic form-factor near zero angle is plotted. In Fig.~\ref{ris:exp_geometry}a one can see additional hexagonally arranged peaks at q of about 10~$\mu$m$^{-1}$. These peaks are actually the cross-sections of the Bragg rods, arising due to the finite thickness of the sample or stacking faults~\cite{chumakova2014periodic}. The intensity of 2$\bar{2}$0, 02$\bar{2}$, 0$\bar{2}$2 and $\bar{2}$20 reflexes should be equal, when the field is applied along [$\bar{1}2\bar{1}$] direction [Fig.~\ref{ris:exp_geometry}b] according to the symmetry, but since the sample was slightly tilted they do not. In Fig.~\ref{ris:exp_geometry}c one can see, that the direct beam has an elliptical form. Moreover the reflexes of 202 family are different by intensity, though should be the same. This is due to the refraction of the beam, since the sample, which represents a thin film inclined to the beam at 54$^{\circ}$ [Fig.~\ref{ris:exp_geometry}i].

\begin{figure}[!htbp]
	\begin{minipage}{0.49\linewidth}
		\center{\includegraphics[width=1\linewidth]{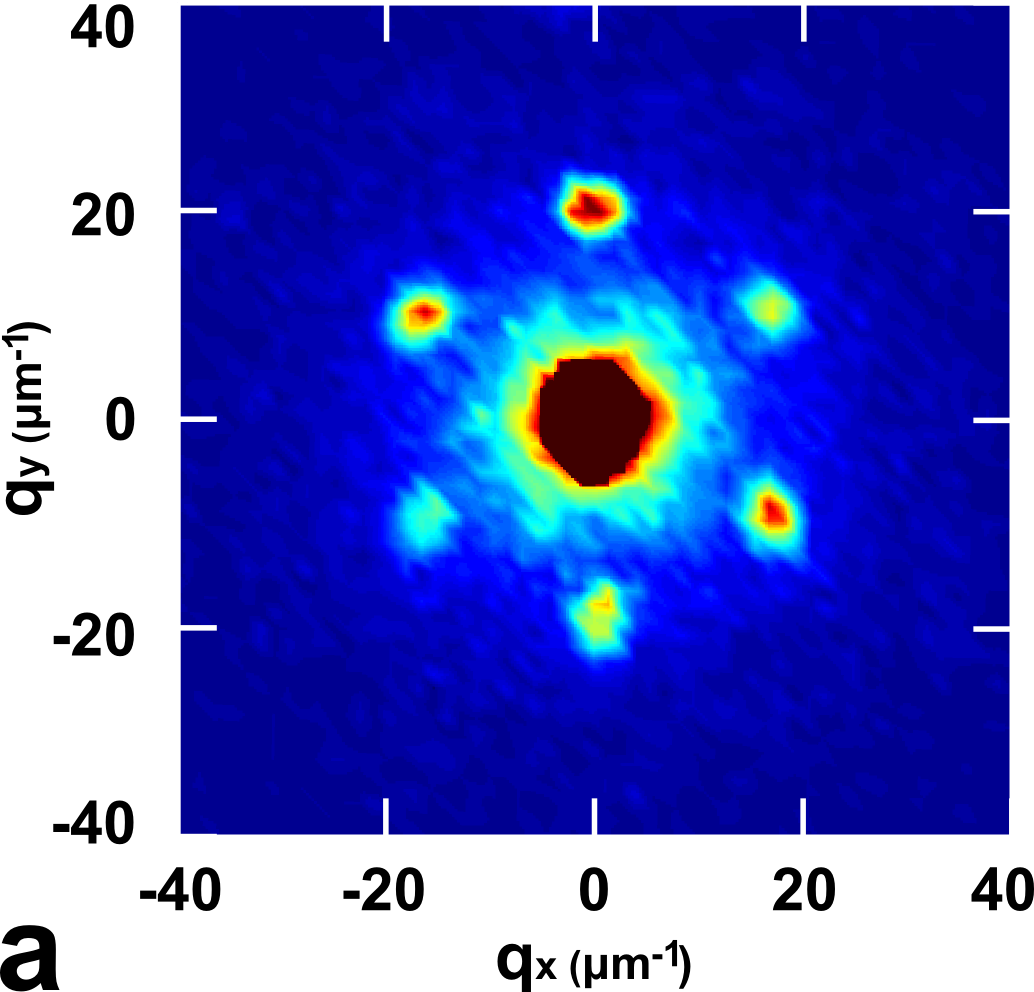}}
    \end{minipage}
	\begin{minipage}{0.49\linewidth}
		\center{\includegraphics[width=1\linewidth]{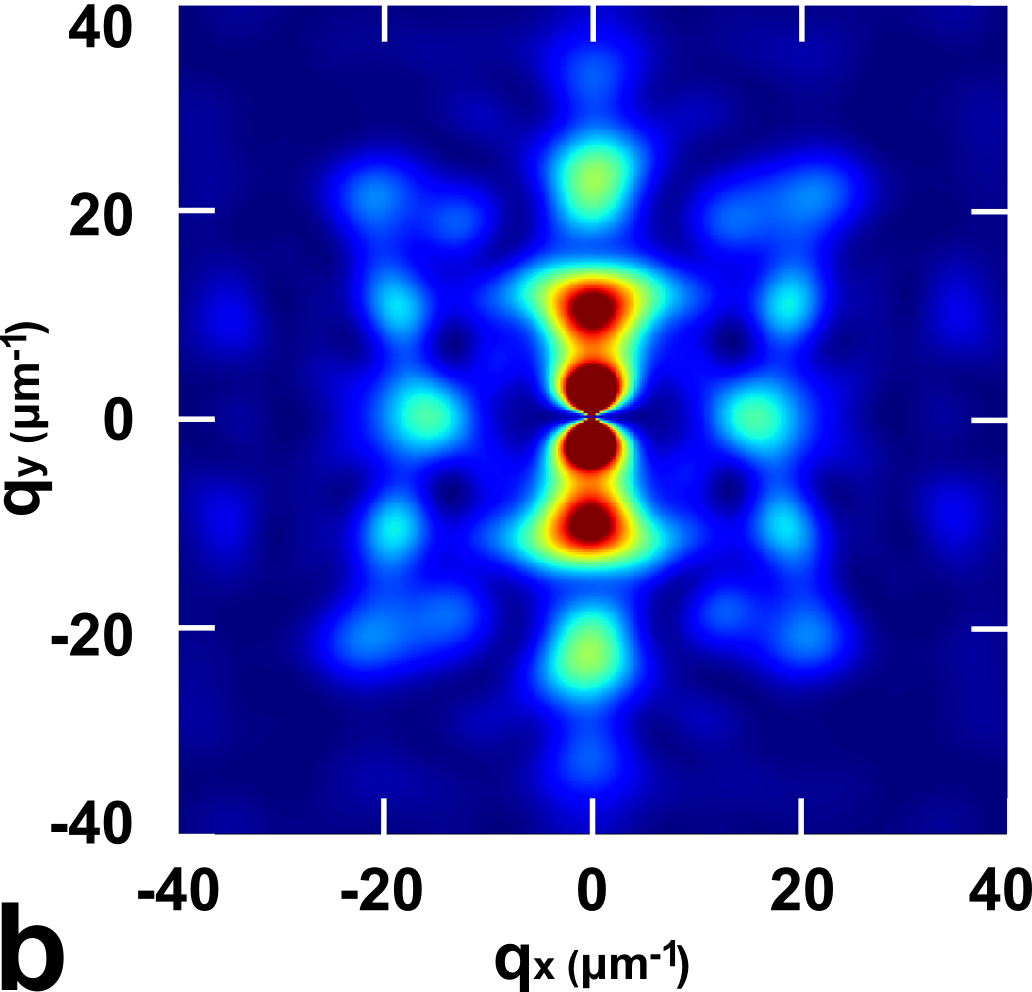}}
	\end{minipage}
	\caption{(Color online) (a) Experimental small-angle neutron diffraction pattern (b) and  corresponding calculated magnetic form-factor in the remanent state, obtained after the external magnetic field \textbf{H} up to 30~kOe was applied along $[\bar{1}2\bar{1}]$ axis and then reduced to zero.}
	\label{ris:rem_2D_pattern}
\end{figure}

It should be noted that in the saturated state the measured and calculated intensities, presented in Fig.~\ref{ris:exp_geometry}, are determined by the spatial structure of the inverse opal. Fig.~\ref{ris:rem_2D_pattern} shows an example of calculated and experimental patterns in the remanent state (the field, was initially applied along $[\bar{1}2\bar{1}]$ axis). The difference is much noticeable as compared with the patterns, obtained in the saturated state.
Similar discrepancy was obtained for the other field directions. This difference is caused by the absence of the magnetic structure factor in calculations. Structure factor minima are expected to suppress significantly the intensity of the observed bright spots of the form factor and \textit{vice versa} its maxima can enhance weak form factor intensity, which manifests itself in the observed intensity pattern. Thus, the calculated form factor distribution does not look as experimental pattern, but nobody should expect it. On the other hand, one can see, that the Bragg peaks in the experimental patterns are situated at the same positions for 30~kOe [Fig.~\ref{ris:exp_geometry}b] as well as for 0~kOe [Fig.~\ref{ris:rem_2D_pattern}a]. Thus, the structural motif of the fcc lattice is caught by the simulation even in the remanent state.

Relatively small coherent volume along with possible defects of the IOLS magnetic structure should lead to minor variation of the structure factor in the external field (i.e. the structure factor for the saturated and remanent state should be virtually the same; see Sec.~\ref{sec:methods}A). Therefore the strongest signal from the magnetic form factor in the Bragg peaks should be contained as the latter is suppressed by the structure factor in other regions of the reciprocal space.

% \hly{[MOVE FURTHER!!!!] This statement is supported by quantitative agreement between simulation and experimental results (see below).}

\begin{figure}[h]
\includegraphics[width=8.5cm]{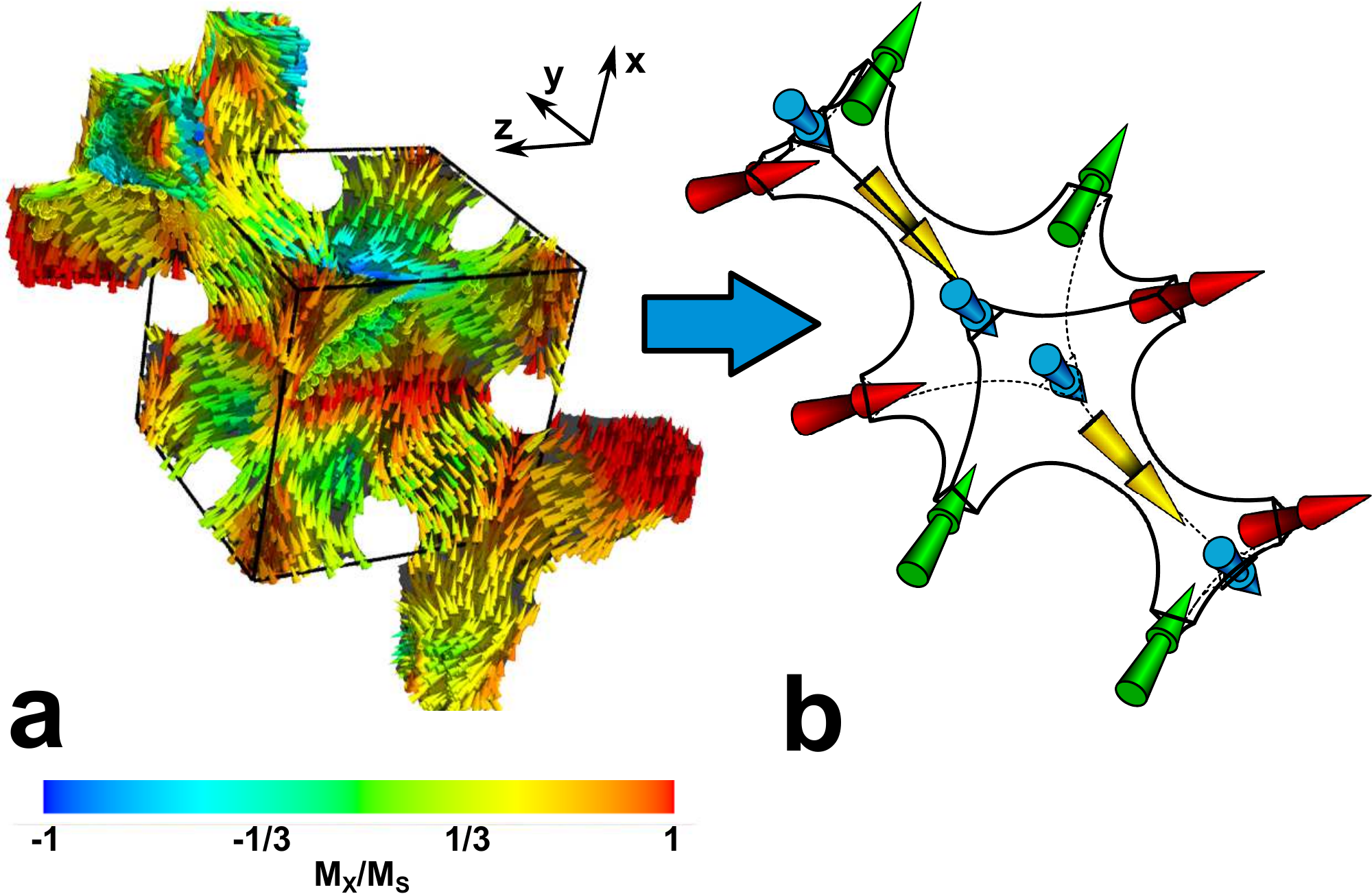}
\caption{(Color online) (a) Calculated magnetization distribution in the IOLS primitive cell. Color indicates the value of x-component of the normalized magnetization. (b) Corresponding simplified scheme. Different colors indicates different $\langle 111\rangle$-type directions.}
\label{ris:prim_cell_calc}
\end{figure}

Since namely the ``legs'' determine magnetic structure of the IOLS, i.e. their reorientation results in strong changes of magnetic state, further we focus on them. The typical magnetization distribution in the IOLS primitive cell is exemplified in Fig.~\ref{ris:prim_cell_calc}a. The calculated SANS intensity depends on the magnetization distribution in the whole unit cell. However for the sake of simplicity we have shown only magnetic state of one primitive cell, it is sufficient for our purposes. Moreover the complicated pictures of the magnetization distribution obtained by micromagnetic simulations were replaced by illustrative schemes, including solely ``legs'' magnetization [Fig.~\ref{ris:prim_cell_calc}b].

With the help of the spin ice model one can easily reveal the state, which the IOLS tends to fall in, if the external field is applied along the given direction. In this state magnetization of the ``legs'' should have a positive projection on the field direction. If the external magnetic field is near the saturation value then the shape anisotropy is suppressed and the ``legs'' magnetization tends to align along the field direction. In the lower fields the ``legs'' magnetization still has a non-negative projection on the field direction but already points along the ``legs''. Further we denote the corresponding magnetic state of the primitive cell as ``target magnetic element'' (TME). Thus, TME is defined by the direction of the external field and corresponds to the configuration of the magnetic moments that is favorable for the field (i.e. leads to the lower Zeeman energy). Depending on the ``convenience'' of TME for the ice rule three scenarios arise.

%%%%%%%%%%%%%%%%%%%%%   [111] SUBSECTION   %%%%%%%%%%%%%%%%%%%%%%%%

\textbf{H~$\parallel$~[111]: Scenario of competition.}
When the field is applied along [111] axis of the IOLS, one ``leg'' in each tetrahedron is parallel to the field direction and the other ``legs'' have equal acute angles to the field --- 72$^{\circ}$ [Fig.~\ref{ris:Basic_element}c].

One can see that the field tries to bring the IOLS into the configuration in which  all ``legs'' magnetic moments have a positive projection to the field direction. However this configuration leads to the ice rule violation i.e. either magnetic charge accumulation (the magnetic moments in three ``legs'' of tetrahedron point inward and the moment of the fourth ``leg'' points outward --- 3-in-1-out state) and sink (3-out-1-in state) arise in tetrahedra. In other words, the field competes with the ice rule. The field decrease should lead to the ice rule fulfillment. Thus, for this geometry, ice rule is obeyed only in the restricted field region, until 3-in-1-out and 1-in-3-out configurations become more profitable in the total energy balance.

Fig.~\ref{ris:H_111} shows the experimental and calculated intensities of 202-type reflexes as a function of the external magnetic field applied along [111] axis. One can see that the curve starts (at H~=~0~kOe) and ends with low value of intensity, i.e. there is no remanent magnetization [Fig.~\ref{ris:H_111}a]. The same results were obtained for hysteresis curves measured by SQUID~\cite{dubitskiy2015spin}. Because of that the determination of the coercivity is challenging. The estimated value is about 0.5~kOe. Hysteresis behaviour can be observed up to 4~kOe, while complete merging of the branches occurs at 6.5~kOe.

The calculations shown, that TME indeed represents a set of 3-in-1-out and 3-out-1-in configurations  (\textbf{B} in Fig.~\ref{ris:H_111}b). However when the field is reduced 2-in-2-out states restore in tetrahedra. Herewith, the range of ice rule fulfillment depends on the ``legs'' shape anisotropy~\cite{dubitskiy2015spin, dubitskiy2017dependence}. The simulation results suggest that the ice rule is valid if field is less than 3.6~kOe (\textbf{C} in Fig.~\ref{ris:H_111}b). In general, the appearance of the calculated curve is very similar to the experimental one. The variation of the intensity and the magnetic form factor from H~=~0 to the coercive field related to fully demagnetized sample (\textbf{A} in  Fig.~\ref{ris:H_111}b) are quite similar. There is also a hysteresis in the low-field part of the calculated curve, but it may represent a double-loop hysteresis.

One can see also that for higher field values the experimental and calculated curves begin to diverge. Therefore real system resists ice rule violation more than the model one. This is possibly caused by demagnetizing field related to the sample shape which may be underestimated in our model.

%%%%%%%%%%%%%%%%%%%%%%%%%

%%%%%%%%%%%%%%%%%%%%%   [1-21] SUBSECTION   %%%%%%%%%%%%%%%%%%%%%%%%

\textbf{H~$\parallel$~[$\bar{1}2\bar{1}$]: Scenario of independence.}

When magnetic field is applied along [$\bar{1}2\bar{1}$] direction two ``legs'' in each tetrahedron are at an angle of $62^{\circ}$ to the field, the third ``leg'' has smaller angle of $19.5^{\circ}$ and the fourth one (along [111]) is perpendicular to the field [Fig.~\ref{ris:Basic_element}c].

TME in this case is defined in the following way: the magnetic moments of the ``legs'', which are inclined to the field, have positive projections on the field direction, while the perpendicular ones due to the ice rule arrange in the same way. Therefore perpendicular to the field component of the magnetization should appear~\cite{mistonov2013three}. It is clear that the direction of the magnetic moments in the ``legs'' that are perpendicular to the field does not affect Zeeman energy. This direction is determined solely by the ice rule. In this sense the ice rule and the field are independent.

The micromagnetic calculations confirm these findings and make it possible to analyze details of magnetization process and interpret SANS data~\cite{dubitskiy2017dependence}. It is convenient to divide all 202-type reflections into two groups: $\textbf{q}_{20\bar2}$  and $\textbf{q}_{\bar202}$ which are perpendicular to the field direction [$\bar{1}2\bar{1}$]; and $\textbf{q}_{02\bar2}$, $\textbf{q}_{0\bar22}$, $\textbf{q}_{2\bar20}$, $\textbf{q}_{\bar220}$ which are at an angle $\alpha$ of $30^{\circ}$ or $150^{\circ}$ to [$\bar{1}2\bar{1}$] direction [Fig.~\ref{ris:exp_geometry}b]. Measured intensity of these reflections groups varies differently in the external field~[Fig.~\ref{ris:H_121}a,c]. The same result has been obtained by means of micromagnetic simulation [Fig.~\ref{ris:H_121}b,d].

The experimental curves [Fig.~\ref{ris:H_121}a,c] start (at H~=~0~Oe) with abrupt decrease of intensity, since the sample was magnetized initially in the direction, which is opposite to the applied field. The minimum at H~=~0.2~Oe corresponds to the coercive force, where the sample should be totally demagnetized. Then the curves, obtained for the peaks of these two groups become different. In Fig.~\ref{ris:H_121}a one can see two-stepped increase of the intensity: rapid growth until 0.45 kOe and smooth gain in higher fields. The saturation field is about 15-20~kOe. The reversal branch repeats the direct one down to 0.45~kOe, where hysteresis arises. At the same time, one can see well-pronounced maximum in the curve in Fig.~\ref{ris:H_121}c. There is no hysteresis after this point, but it takes place in smaller fields. Interestingly, that the intensity drops abruptly in the zero field point.

Maximum value of the calculated intensity shown in Fig.~\ref{ris:H_121}d  is reached when magnetization in the ``legs'', which are inclined to the field, has a positive projection on the field direction, while magnetization of the ``legs'' those are perpendicular to the field is oriented in the same way (i.e. magnetic configuration corresponds to TME) (state \textbf{C}). Further field increase leads to the inclination of the magnetization in  all ``legs'' towards the field direction. Thus, the value of $\widetilde{\textbf{M}}_{\perp}(\textbf{q})$ decreases and hence peak intensity also decreases (see Eq.~\ref{eq:magnetic_section}).

If one starts to decrease the external field then the magnetic moments in the ``legs'' which are perpendicular to the field reverse in order to minimise the whole demagnetization energy of the system. As a result perpendicular to the field magnetization component decreases. A rapid drop of this component takes place till the field of  0.6~kOe (transition from state \textbf{C} to \textbf{D}). Decreasing branch of calculated curve exhibits a kink in this field. The same but less pronounced kink is present in the experimental curve in the field of 0.4~kOe. Minimum of the intensity corresponds to fully demagnetized state (state \textbf{A}).

It is worth noting that magnetic moments of the ``legs'' which are at minimum angle to the external field (red in the Fig.~\ref{ris:H_121} reverse in smaller field range than perpendicular ones (state \textbf{B}). This range coincides with the hysteresis region range in Fig.~\ref{ris:H_121}b. Reflections of the $\textbf{q}_{20\bar2}$ group are more sensitive to the magnetization of these ``legs'' than $\textbf{q}_{02\bar2}$ ones.

One can conclude that remagnetization process starts with reversal of magnetic moments in the ``legs'' perpendicular to the field. Due to the ice rule this reversal is accompanied by tilt of the magnetic moments in ``legs'' that are at angle of $62^{\circ}$ to the field. After that the magnetization of the ``legs'' that are at angle of $19.5^{\circ}$ reverses. This model is supported by quantitative agreement between the calculated and experimental hysteresis curves.

%%%%%%%%%%%%%%%%%%%%%   [100] SUBSECTION   %%%%%%%%%%%%%%%%%%%%%%%%

\textbf{H~$\parallel$~[100]: Scenario of cooperation.}
When the field is applied along [100] direction all angles between the ``legs'' and the field are the same and equal to $54.7^{\circ}$ [Fig.~\ref{ris:Basic_element}c]. In this case the state with positive projections of the magnetic moments on the field direction (TME) automatically obey the ice rule. In other words the ice rule and the field ``cooperate'' to minimize the total energy.

The experimental and calculated intensities of 202-type reflexes as a function of external magnetic field applied along [100] direction are shown in Fig.~\ref{ris:H_100}. One can see, that the look of the experimental curve [Fig.~\ref{ris:H_100}a] is similar to one, presented in Fig.~\ref{ris:H_121}a. It has initial decrease of intensity, while field increasing from 0~Oe to the coercivity at 0.2~Oe, abrupt and smooth increase regions and decrease of intensity after saturation. Even the values of the critical fields are very close to each other. The calculated curve [Fig.~\ref{ris:H_100}b] has the same form, but abrupt increase of intensity is two-stepped and decrease of intensity to the final state is also abrupt in contrast to the experimental curve. These jumps of the calculated curve are caused by ``legs'' magnetic moments reversals. In the real system they are less pronounced most likely because of large number of ``legs'' and spread of their shape anisotropy.

Because of the ``cooperation'', during field decrease till the field of 0.4~kOe state \textbf{C} (TME) is stable. However one can see the fall of the calculated Fourier transform in Fig.~\ref{ris:H_100}(b). The same fall was obtained in the experiments (Fig.~\ref{ris:H_100}(a)). This drop can not be interpreted in terms of the spin ice model accounting for the reorientation of magnetic moments in the ``legs''. The reason is discussed below in ``Beyond spin ice model'' subsection. Further field decrease leads to ``legs'' magnetization reversal and substantial drop of the form-factor value at 0~kOe (state \textbf{D}). Then IOLS falls in fully demagnetized state (state \textbf{A}). However the magnetization restores rapidly during field growth by two-step process. Firstly the system reaches state \textbf{B} and then in the field of 1.2~kOe the magnetic moments in all ``legs'' complete reversal (state \textbf{C}). The branches of the hysteresis curve merge only when the field value reaches 1.8~kOe. This discrepancy also can not be explained by the ``legs'' magnetization reversal model.

For all considered field directions the branches of hysteresis loops merge when the magnetic moments in the ``legs'' stop to reverse and fall into the final state. According to the spin ice model the branches should merge at the lowest field value if the ice rule and magnetic field ``cooperate'' and at the highest one otherwise. Both the experimental and calculated results seem to be in agreement with this model. The merging point was found to be 1~kOe, 3.5~kOe and 7~kOe when the field is applied along [100] (cooperation), [$\bar{1}2\bar{1}$] (independence), [111] (competition) directions respectively. the corresponding calculated values are 1.8~kOe (field along [100]), 3~kOe ([$\bar{1}2\bar{1}$]) and 6~kOe ([111]).

To make a tentative conclusion, the experimentally obtained hierarchy of the hysteresis loop merging points is indeed linked to the interplay between the magnetic field and the ice rule. Nevertheless, a rigorous interpretation of the experimental results is not possible in the frame of the spin ice model. Nonuniform states arisen in the IOLS structural elements have to be taken into account by means of the micromagnetic calculations. More detailed discussion is presented in the next subsection.

%%%%%%%%%%%%%%%%%%%%%%%%%%%% BEYOND THE MODEL %%%%%%%%%%%%%%%%%%%%%%%%%%%

\textbf{Beyond the spin ice model.}
The spin ice model allows one to explain main features of the experimental results. However, there are two important factors, which are not taken into account by this approach.

\begin{figure*}[!htbp]

  \includegraphics[width=0.99\linewidth]{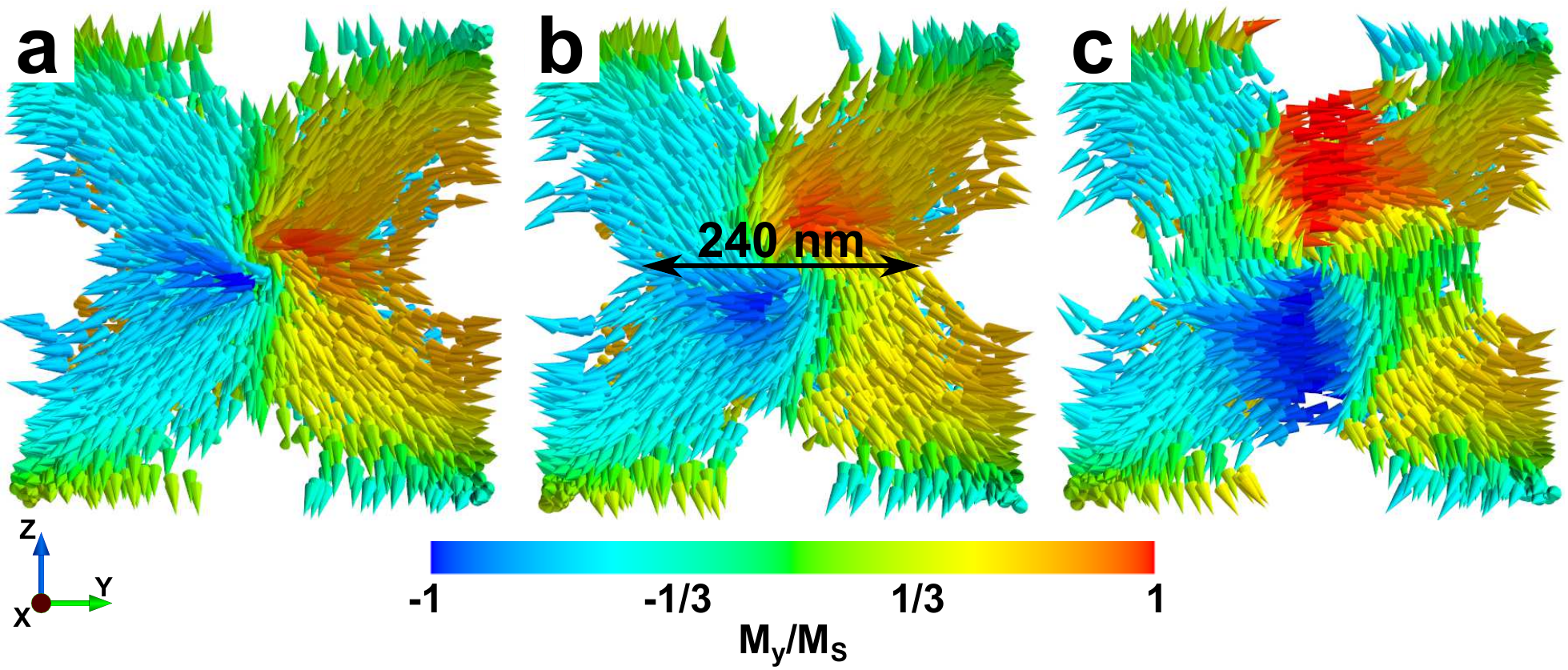}
  \caption{(Color online) Vortex state evolution in a cube when the field is applied along [100] axis: (a) 4~kOe, (b) 1.6~kOe, (c) 0.2~kOe, descending branch of the magnetization reversal curve}
  \label{ris:vortex_100}

\end{figure*}

The first one is the shape anisotropy caused by the entire sample shape. Since the sample represents a thin film (12~$\mu$m) the strongest stray field appears, when the external field is perpendicular to the sample plane (along [111] axis), and the weakest one~---~when the field is in the sample plane (along [$\bar{1}2\bar{1}$] axis). Intermediate case is realized when the external field is applied along [100] axis. In the calculations presented here, the demagnetization effects were taken into account, that significantly improved the agreement between the simulated and experimental curves.

It is worth noting, that demagnetization could change the hierarchy of the branches merging fields observed in different geometries. However, nor in the experiment, neither in the calculations this hierarchy was not altered.

Secondly the magnetic moments reversal in the ``legs'' is generally accompanied by abrupt changes of the magnetic form factor. However the calculated form factor field dependencies also contain regions of smooth form factor variation.

In particular state \textbf{C} in all geometries is stable in a relatively large field range during the field decrease. In this state the spin ice rule is fulfilled therefore the external magnetic field competes solely with the ``legs'' shape anisotropy. Nevertheless one can notice reduction of both the measured intensity and calculated magnetization Fourier transform when the field decreases although the ``legs'' configuration is not changed. It was found that this reduction is caused by the magnetization rotation in tetrahedra and particularly by the vortex states arisen in the cube. The reason why the vortex states are likely to appear in the cube is that its linear size is relatively large (approximately 240~nm). The vortices were observed for all considered field directions.

The most considerable influence of the vortex on the form factor field dependence was found when the field was applied along [100] axis (``cooperation'' case). The vortex evolution causes form factor variation up to the field of 0.4~kOe in the case of the decreasing branch of the hysteresis loop and from the field of 1.2~kOe in the case of increasing one (state \textbf{C}). Additionally, upon field increase vortex state becomes reversible only in the field of 1.8~kOe. One can suggest that in the experiment all the magnetic moments in the ``legs'' fall in the final state at 1~kOe and additional hysteresis loop between 1~kOe and 2.5~kOe is caused by non-uniform states in the cube. Snapshots of the vortex evolution are shown in Fig.~\ref{ris:vortex_100}. Thus not only the anisotropic ``legs'' control the magnetic properties of the IOLS, but also non-uniform magnetic states. Therefore inverse opals have additional degrees of freedom compared to the conventional atomic spin-ices.
In general, not complete agreement between the experimental and calculations can be linked to the bunch of factors: 1) local defects of IOLS structure which lead to spread in ``legs'' reversal fields values; 2) defects of IOLS mesostructure such as twinning and stacking faults; 3) magnetocrystalline anisotropy of cobalt grains; 4) partial account of magnetization field related to the whole sample shape; 5) exclusion of the structure factor variation from our model. Besides that, some experimental imperfections (e.g. inaccuracy of the sample alignment in the beam) can give rise to additional inconsistence. However based on the comparison of the experimental and simulated data we suggest that the magnetic structure of three-dimensional artificial mesoscaled spin ice was resolved.

~\\

%%%%%%%%%%%%%%%%%%%%%   CONCLUSION SECTION   %%%%%%%%%%%%%%%%%%%%%%%%

\section{Concluding remarks}
\label{sec:conclusion}
We successfully uncovered magnetic states of the inverse opal-like structure during the magnetizing process by using complementary small-angle neutron diffraction and micromagnetic calculations. Although we have considered only specific points of the magnetization curves, one can also reveal magnetic state in the intermediate points. Owing to the micromagnetic simulations it is possible to find magnetization distribution, while by using experimental data, one can adjust field values corresponded to this distribution.

According to the performed calculations, in the geometry, when the field is along [111] axis 3-in-1-out and 1-in-3-out configurations indeed arise in the tetrahedra. When the field is along [$\bar{1}2\bar{1}$] axis the ice rule gives rise to perpendicular to the field component of magnetization. When the field is along [100] axis system magnetizes in the fastest way, despite larger demagnetizing field, than in the case, when H is along [$\bar{1}2\bar{1}$]. The ice rule has the range of validity for the first case, while for two others it is fulfilled up to high fields, until shape anisotropy of the ``legs'' holds magnetization.

We have shown, that two-dimensional patterns of magnetic form-factor contribution are in good agreement with experimental diffraction maps taken at saturation state for all considered external field directions. But at low field values, the calculated and experimental results differ. Nevertheless calculated intensity dependencies of Bragg peaks on the external field are in semi-quantitative agreement with the experimental ones.

We speculate, that the suggested approach can be successfully applied for the revealing the magnetic structure of complicated three-dimensional periodic systems.

%%%%%%%%%%%%%%%%%%%%%   ACKNOWLEDGEMENTS SECTION   %%%%%%%%%%%%%%%%%%%%%%%%

\section{Acknowledgements}
Authors thank A.V. Syromyatnikov for the fruitful discussions. Also authors are grateful to the staff of SANS-1 beamline at FRM-II, especially to S.-A.~Siegfried for the hospitality and invaluable help during experimental time.
Scanning electron microscopy studies were performed at the Research park of St.Petersburg State University Interdisciplinary Center for Nanotechnology (http://nano.spbu.ru/index.php/en) and micromagnetic calculations were performed using computational resources provided by Resource Center ``Computer Center of SPbU'' (http://www.cc.spbu.ru/en). Work was supported by Russian Foundation for Basic Research (pr.~ofi-m 14-22-01113).
%
%\section{Acknowledgements}
% Authors acknowledge N.A.~Sapoletova, K.S.~Napolskii and A.A.~Eliseev from the Department of Materials Science of Moscow State University for samples synthesis. We thank to Interdisciplinary Center for Nanotechnology of Research park of St.~Petersburg State University for the SEM investigations. The work is in part by the Russian Foundation for Basic Research (Project No. 14-22-01113).
%
%
\bibliography{biblio}

\end{document}